\documentclass[aps, pre, twocolumn,floatfix,superscriptaddress, 10pt]{revtex4-2}

\usepackage{mathtools}
\usepackage{amsmath}
\usepackage{breqn}
\usepackage{graphicx}
\usepackage{textcomp, gensymb}

\begingroup\makeatletter
\catcode`,=\active
\global\let\breqn@comma,
\protected\gdef,{\ifmmode\expandafter\breqn@comma\else\expandafter\active@comma\fi}
\endgroup

\begin{document}

\title{Preserving elastic anisotropy with tessellations of granular packings}

\author{Annie Z. Xia}
\affiliation{Department of Mechanical Engineering, Yale University, New Haven, Connecticut 06520, USA}
\affiliation{Integrated Graduate Program in Physical and Engineering Biology, Yale University, New Haven, Connecticut 06520, USA}

\author{Dong Wang}
\affiliation{Department of Mechanical Engineering, Yale University, New Haven, Connecticut 06520, USA}

\author{Catherine La Riviere}
\affiliation{Department of Physics, University of Michigan, Ann Arbor, Michigan 48109}

\author{Rebecca Kramer-Bottiglio}
\affiliation{Department of Mechanical Engineering, Yale University, New Haven, Connecticut 06520, USA}

\author{Mark D. Shattuck}
\affiliation{Benjamin Levich Institute and Physics Department, The City College of The City University of New York, New York, New York 10031, USA}

\author{Corey S. O'Hern}
\affiliation{Department of Mechanical Engineering, Yale University, New Haven, Connecticut 06520, USA}
\affiliation{Integrated Graduate Program in Physical and Engineering Biology, Yale University, New Haven, Connecticut 06520, USA}
\affiliation{Department of Physics, Yale University, New Haven, Connecticut 06520, USA}
\affiliation{Department of Applied Physics, Yale University, New Haven, Connecticut 06520, USA}

\begin{abstract}

Multiscale periodic metamaterials have been designed for numerous applications, such as impact absorption, acoustic cloaking, photonic band gaps, and mechanical logic gates. This prior work has focused on optimizing mesoscale structure for desired bulk isotropic properties. In contrast, we seek to develop materials with highly anisotropic elastic properties. To quantify elastic anisotropy, we introduce two rotationally invariant, normalized quantities that characterize the anisotropic response to shear and compression, respectively, $A_G$ and $A_C$. We find that typical crystalline solids possess average elastic anisotropy $\overline{A}_G \approx 0.15$ and $\overline{A}_C \approx 0.09$. Compared to atomic crystals, jammed granular materials can attain elastic anisotropies that are several orders of magnitude larger. Since grain rearrangements reduce anisotropy in granular materials, to preserve strong elastic anisotropy, we design tessellated granular materials that consist of multiple connected grain-filled voxels, which limit rearrangements and enable highly anisotropic elastic properties. Bulk granular packings with $N$ grains prepared at pressure $p$ have maximal anisotropy for $pN^2\sim1$ and become isotropic in the large-$pN^2$ limit. We show that homogeneously tessellated granular systems can inherit the elastic response of the constituent voxel configurations with elastic anisotropy up to $100$ times that of crystalline compounds over a range of $pN^2$. We show further methods to tune the elastic anisotropy of tessellations by designing heterogeneously patterned voxel configurations and tessellations that allow large boundary deformations. 

\end{abstract}

\maketitle

\section{Introduction}
Previous studies have designed periodic metamaterials, e.g. lattice-structured or architectured materials~\cite{liu_mechanical_2019, lee_micronanostructured_2012, rosa_enhanced_2025, zheng_multiscale_2016, benedetti_architected_2021}, for applications in acoustic cloaking, phononic band 
gaps~\cite{mousanezhad_honeycomb_2015,chen_periodic_2014, chen_acoustic_2007, yang_acoustic_2010}, impact absorbers~\cite{yuan_3dprinted_2019, di_frisco_structural_2024, jiang_elastically_2023}, auxetic response~\cite{javid_dimpled_2015, greaves_poissons_2011, liu_mechanics_2018, gao_two-dimensional_2018}, 
high tensile strength at low mass density~\cite{seto_tough_2008, keaveny_biomechanics_2001, moestopo_pushing_2020, bauer_approaching_2016, berger_mechanical_2017, filipov_origami_2015}, and mechanical computation~\cite{yasuda_mechanical_2021, song_additively_2019, yasuda_origami-based_2017, treml_origami_2018, ion_digital_2017}. This prior work has primarily focused on designing the material's isotropic elastic response. In contrast, we seek to design periodic structures with a strongly anisotropic elastic response, e.g. compliant in one direction and stiff in another. An isotropic elastic material has two independent elastic moduli that remain the same when measured in any coordinate frame. In contrast, an anisotropic elastic material in three dimensions can have up to $21$ independent elastic moduli $C_{mn}$. Current measures of elastic anisotropy include the Zener ratio $C_{44}/(C_{11}-C_{12})$ and other ratios and differences among the elements of the elastic modulus tensor. However, these quantities are not rotationally invariant and can vary significantly with the choice of the coordinate system. In this work, we focus on two simple rotationally invariant measures of the anisotropic elastic response, $A_G$ and $A_C$, which capture the normalized variance in the angle-dependent shear and compressive response~\cite{goodrich_jamming_2014}. 
We first evaluate the elastic anisotropy for a database of 1181 crystalline solids with elastic moduli calculated using density functional theory (DFT)~\cite{de_jong_charting_2015}. We find that these crystalline materials possess relatively small values for the elastic anisotropy: $\overline{A}_G=0.15\pm0.09$ and $\overline{A}_C=0.09 \pm 0.08$. 

Due to their amorphous structure, jammed granular packings are promising candidates as materials with strongly anisotropic elastic response~\cite{jaeger_granular_1996, liu_jamming_2010, behringer_physics_2018}. In this work, we show that granular packings can possess elastic anisotropy that is two orders of magnitude larger than that for atomic crystals. The ensemble-averaged $\overline{A}_G$ and ${\overline A}_C$ for granular packings collapse as a function of $pN^2$ and reach a maximum near $pN^2\sim1$. In addition, the elastic response of granular packings depends sensitively on their intergrain contact networks, which can change as the packings deform during applied loads~\cite{tsamados_local_2009, manning_vibrational_2011, cubuk_identifying_2015, falk_dynamics_1998, zaccone_approximate_2011}. On average, grain rearrangements give rise to more isotropic elastic properties in granular packings~\cite{goodrich_jamming_2014, zhang_local_2023}. The likelihood for grain rearrangements increases with system size so larger granular packings more frequently rearrange~\cite{shang_elastic_2020, maloney_subextensive_2004}. In contrast, as plasticity in crystals is localized to topological defects, the elastic response of crystalline solids does not possess significant system-size dependence ~\cite{taylor_mechanism_1934}. To enhance the elastic anisotropy of granular solids, as well as minimize homogenizing grain rearrangements, we investigate tessellated granular materials in three dimensions (3D).

Tessellated granular materials consist of an array of connected voxels that each contain a small number of spherical grains ($N\leq16$) confined by physical boundaries. Previous work has shown that 2D tessellated granular materials successfully limit grain rearrangements, ensuring a decreasing shear modulus as a function of increasing pressure~\cite{pashine_tessellated_2023, zhang_designing_2023}. In this work, we investigate 3D homogeneous tessellations of the same granular packing in each voxel, which can inherit the anisotropic elastic properties of the individual granular packings. We also study 3D heterogeneous tessellations, which consist of two or more different grain configurations within the tessellations. We find that $A_G$ and $A_C$ of 3D heterogeneous tessellations can be larger than the average response of individual voxels by up to an order of magnitude, and can be tuned by controlling the non-affine deformation of the tessellation boundaries. These results emphasize that designing tessellations of voxels containing granular packings is a promising way to maximize the elastic anisotropy of bulk materials.

This article is organized as follows. In Sec.~\ref{sec:methods}, we discuss the numerical methods employed to generate jammed granular packings, calculate their elastic moduli, and quantify their elastic anisotropy using periodic boundaries, single voxels, and tessellations of multiple voxels. In Sec.~\ref{sec:results}, we present the main results. We show that the compressive $A_C$ and shear $A_G$ elastic anisotropy are significantly larger in granular packings than in atomic crystals and elastic spring networks derived from the atomic crystals. We also verify that the material symmetry encoded in the elastic modulus tensor $C_{mn}$ influences $A_C$ and $A_G$, i.e. small values of elastic anisotropy occur when $C_{mn}$ is symmetric, whereas large values of elastic anisotropy occur when $C_{mn}$ is strongly asymmetric. Finally, we show that tessellated granular materials minimize grain rearrangements, which preserves the large elastic anisotropy of granular packings with $A_C$ and $A_C$ up to $10^2$ times larger than the values for atomic crystals. In Sec.~\ref{sec:conclusions}, we discuss the conclusions and propose promising directions for future research, including employing machine learning methods to explore the large space of heterogeneous tessellations to maximize anisotropic elastic properties. In addition, the article includes six appendices. In Appendix \ref{si:box_potential} we write the flexible boundary potential in terms of the box vertex positions ${\vec b}_i$. In Appendix \ref{si:strains}, we specify the six strain tensors required to calculate $C_{mn}$. In Appendix \ref{si:au}, we discuss other measures of elastic anisotropy and show that another commonly used measure of elastic anisotropy, $A^U$, is similar to $A_G$ and $A_C$. Appendix~\ref{si:spring_nets} details the construction of the linear spring networks derived from crystalline compounds~\cite{de_jong_charting_2015}. In Appendix \ref{si:stability}, we describe the protocol to generate random elastic modulus tensors $C_{mn}$ with varying symmetry strength and compare the mechanical stability of randomly generated $C_{mn}$ to that of granular packings. In Appendix \ref{si:analytics}, we provide analytical expressions for the elements of $C_{mn}$ that define $A_C$ and $A_G$.
\section{Methods}
\label{sec:methods}
The Methods section contains three subsections. In Sec.~\ref{methods_a:jamming}, we discuss the grain-grain interactions, grain-wall interactions, and internal wall interactions, the different boundary conditions studied, and the numerical procedure used to generate jammed granular packings. In Sec.~\ref{methods_b:c_tensor}, we introduce the numerical methods used to calculate the elastic modulus tensor $C_{mn}$ for granular packings and tessellations. In Sec.~\ref{methods_c:anisotropy}, we describe the measures of elastic anisotropy used in this work, which are coordinate invariant functions of $C_{mn}$.

\begin{figure}
\centering
\includegraphics[width = 1.0\columnwidth]{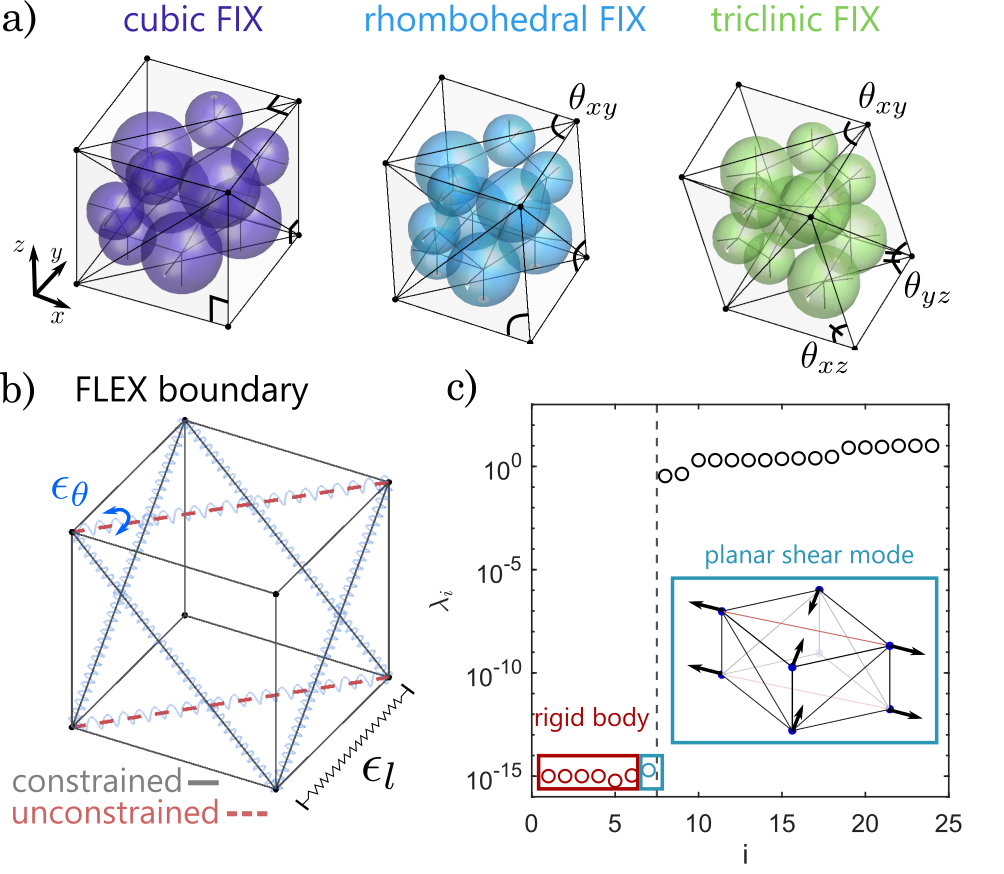}
\caption{(a) Types of fixed boundary conditions (FIX) for granular packings: cubic ($\theta_{xy}=\theta_{xz}=\theta_{yz}=90\degree$), rhombohedral ($\theta_{xy}=\theta_{xz}=\theta_{yz}=75\degree$), and triclinic ($\theta_{xy}=75^{\circ}$, $\theta_{xz}=60^{\circ}$, and $\theta_{yz}=45\degree$), where $\theta_{ij}$ is the angle in the $ij$ plane between boundary edges.
(b) Flexible boundary conditions (FLEX) with $6$ faces that have non-zero bending energy $\epsilon_\theta$ (blue helix) between the two triangles in each face and $16$ edges with non-zero edge length energy $\epsilon_l$ (black solid lines). The two diagonal dashed lines indicate edge length constraints with $\epsilon_l=0$ (red dashed lines).
(c) Eigenvalues $\lambda_i$ of the dynamical matrix $M$ of the voxel with FLEX boundary conditions in (b) without grains, plotted versus the index $i$ of the sorted eigenvalues. The empty voxel has six rigid-body zero eigenvalues (red circles) and one planar shear mode in the $x$-$y$ plane with $\lambda_i=0$ (light blue circle). The inset shows deformation of the empty voxel along the shear mode.}
\label{fig:methods}
\end{figure}
\subsection{Grain interactions and protocol to generate jammed granular packings}
\label{methods_a:jamming}

We study static packings of $N$ bidisperse frictionless, purely repulsive soft spheres, with $N/2$ small and $N/2$ large spheres and diameter ratio $\sigma_l / \sigma_s = 1.4$. 
We consider three types of boundary conditions for the granular packings: periodic boundary conditions and two types of physical boundary conditions. The spheres interact via the following pairwise, purely repulisve interaction potential: 
\begin{equation}
U_{gg}(r_{ij}) = \frac{\epsilon_{gg}}{2} \left( 1-\frac{r_{ij}}{\sigma_{ij}} \right) ^2 \, \Theta \left( 1-\frac{r_{ij}}{\sigma_{ij}} \right),
\label{gg}
\end{equation}
\noindent where $\epsilon_{gg}$ is the strength of the repulsive interactions, $r_{ij}$ is the separation between the centers of grains $i$ and $j$, $\sigma_{ij}$ is the sum of the radii of grains $i$ and $j$, and $\Theta(\cdot)$ is the Heaviside step function, which ensures that the grains do not interact when they are not in contact, $r_{ij} > \sigma_{ij}$. For periodic boundary conditions (PBC), the central simulation cell is surrounded by $26$ nearest-neighbor image cells that are identical to the central cell and the total potential energy of the system is $U = \sum_{i>j=1}^N U_{gg}(r_{ij})$, which is only a function of the grain separations.

For physical boundaries, we include repulsive interactions between the grains and boundary surfaces using the purely repulsive linear spring potential similar to that in Eq.~\ref{gg}:
\begin{equation}
U_{gw} \left( r_{ik} \right) = \frac{\epsilon_{gw}}{2} \left( 1-\frac{r_{ik}}{\sigma_{i}} \right)^2 \, \Theta \left( 1-\frac{r_{ik}}{\sigma_{i}} \right),
\end{equation}
\noindent where $\epsilon_{gw}$ is the strength of the repulsive interaction, $r_{ik} = \left| ({\vec r}_i - {\vec b}_{k}) \cdot {\hat n}_{k} \right|$ is the distance between the center of grain $i$ and triangular wall face $k$ along the normal direction ${\hat n}_k$, where ${\vec r}_i$ is the position of the center of grain $i$ and ${\vec b}_k$ is the position of any vertex on triangular wall face $k$. 

We will also consider deformations of the physical boundaries using the following shape-energy function:  
\begin{multline}
    U_w = \frac{\epsilon_v}{2} \left( \frac{v}{v_0} - 1 \right)^2 + \epsilon_a\sum_{i=1}^{N_f} \left( \frac{a_i}{a_{i,0}} - 1 \right)^2 + \\
    \epsilon_l\sum_{j=1}^{N_e} \left( \frac{l_j}{l_{j,0}} - 1 \right)^2 + \epsilon_{\theta}\sum_{k=1}^{N_f/2} \left( \frac{\theta_k}{\theta_{k,0}} - 1 \right)^2, 
\label{eqn:uw}
\end{multline}
which includes quadratic energy penalties when the boundary edge lengths, triangle face areas, volume, and bending angles between triangle faces deviate from their preferred values: $l_{j, 0}$, $a_{i, 0}$, $v_0$, and $\theta_{k,0}$, respectively. In Eq.~\ref{eqn:uw}, $l_j$, $a_i$, $v$, and $\theta_k$ are the current edge lengths, triangle face areas, volume, and bending angles of the voxel boundary~\cite{wang_structural_2021}, which can be written in terms of the vertex positions ${\vec b}_i$ as shown in Appendix \ref{si:box_potential}. The strength of the edge length, triangle area, volume, and bending energies are controlled by $\epsilon_l$, $\epsilon_a$, $\epsilon_v$, and $\epsilon_\theta$. 
We set the units for length, energy, and stress as $\sigma_s$, $\epsilon_{gg}$, and $\epsilon_{gg}/\sigma_{s}^3$. The results below are shown for $\epsilon_{gw}/\epsilon_{gg}=\epsilon_v/\epsilon_{gg}=\epsilon_a/\epsilon_{gg}=\epsilon_l/\epsilon_{gg}=\epsilon_\theta/\epsilon_{gg}=1$. 

For the first physical boundary condition, the positions of the boundary are held fixed (FIX; Fig.~\ref{fig:methods} (a)), while for the other physical boundary condition the boundary vertex positions are evolved according to the grain-boundary $U_{gw}$ and boundary shape $U_w$ energy functions (FLEX; Fig.~\ref{fig:methods} (b)). For both the FIX and FLEX boundaries, we study cubic voxels with the minimal number of vertices $N_v=8$, triangular faces $N_f=12$, and edges $N_e = 18$ needed to triangulate their surfaces. We focus on parallelepiped boxes, which are the simplest non-cubic 3D shape able to tessellate space. For the FIX boundary condition, we study packings jammed within different fixed parallelepipeds, described by the angles $\theta_{xy}, \theta_{xz}, \theta_{yz}$ that determine the in-plane angles of each face. A parallepiped with three different angles $\theta_{xy}$, $\theta_{xz}$, and $\theta_{yz}$ corresponds to the unit cell of a crystal with triclinic symmetry. We also study parallelepipeds that are more symmetric, such as the rhombohedral unit cell where $\theta_{xy}=\theta_{xz}=\theta_{yz}$ and cubic unit cell with $\theta_{xy}=\theta_{xz}=\theta_{yz}=90\degree$. The total potential energy of a jammed packing with the FIX boundary condition is the sum of contributions from grain-grain and grain-wall interactions: $U=\sum_{i>j=1}^N U_{gg}(r_{ij})+\sum_{i=1}^{N}\sum_{k=1}^{N_f} U_{gw}(r_{ik})$.

FLEX boundaries are constructed so that an empty voxel can deform with zero energy cost along a single shear plane, which allows the voxels to be compliant prior to adding the grains. (See Fig.~\ref{fig:methods} (c).) To construct FLEX boundaries, we start with all $N_e=18$ edges connecting pairs of vertices with $\epsilon_l>0$ and no area or volume constraints ($\epsilon_{a}=0$ and $\epsilon_v=0$). Then, we remove the two diagonal edge length constraints in either the $x$-$y$, $x$-$z$, or $y$-$z$ plane to yield $16$ edge length constraints. We enforce $N_{\textrm{p}}=6$ bending constraints $\epsilon_\theta >0$ between adjacent triangular faces to prevent out of plane bending. Enforcing these constraints on an empty voxel gives rise to one nontrivial zero eigenvalue $\lambda_i$ out of the $3N_v=24$ eigenvalues of the dynamical matrix for an empty voxel: 
\begin{equation}
M_{i_\alpha, j_\beta} = \frac{\partial^2 U_w}{\partial b_{i, \alpha} \partial b_{j,\beta}},
\end{equation}
where $\alpha$ and $\beta$ refer to the Cartesian components of the vector $\vec{b_i}$ which is the position of vertex $i$. In Fig.~\ref{fig:methods} (c), we show the sorted $\lambda_i$ to verify that there are $6$ rigid-body zero eigenvalues and one nontrivial zero eigenvalue corresponding to shear in the $x$-$y$ plane. The total energy of a jammed packing with the FLEX boundary condition is $U=\sum_{i>j=1}^N U_{gg}(r_{ij})+\sum_{i=1}^{N}\sum_{k=1}^{N_f} U_{gw}(r_{ik}) + U_{w}({\vec b}_i)$. We define the force ${\vec f}_i$ on grain $i$ as the gradient of the potential energy with respect to the grain position,
\begin{equation}
{\vec f}_{i} = -\partial U / \partial {\vec r}_{i}.
\end{equation}
Similarly, for FLEX boundary conditions, the force ${\vec f}^b_i$ on boundary vertex $i$ is
\begin{equation}
{\vec f}_{i}^b = -\partial U / \partial {\vec b}_{i}.
\end{equation}
We calculate the stress tensor $\Sigma_{\alpha \beta}$ using the virial expression, where $\alpha$ and $\beta$ index $x$, $y$, and $z$ as follows:
\begin{equation}
\label{eqn: virial}
   \Sigma_{\alpha \beta} = \frac{1}{V}\left(\sum_{i=1}^{N} f_{i, \alpha} r_{i, \beta}+\sum_{j=1}^{N_v} f_{j, \alpha}^b b_{j, \beta}\right),
\end{equation}
where $V$ is the total volume of the system, $f_{i,\alpha}$ is the $\alpha$-component of the total force on grain $i$, $r_{i,\beta}$ is the $\beta$-component of the position vector of grain $i$, $f_{j,\alpha}^b$ is the $\alpha$-component of the total force on boundary vertex $j$, and $b_{j,\beta}$ is the $\beta$-component of the position vector of boundary vertex $j$. The total pressure of the system is defined as
\begin{equation}
\label{eqn: pressure}
p = \frac{1}{3} \left(\Sigma_{xx} + \Sigma_{yy} + \Sigma_{zz} \right).
\end{equation}
To generate jammed granular packings (for all three types of boundary conditions), we first randomly place grains at a dilute packing fraction of $\phi < 0.01$. We then grow the grains in small packing fraction increments $\Delta \phi / \phi= 10^{-3}$, and minimize the total potential energy after each packing fraction increment using the Fast Inertial Relaxation Engine (FIRE) method until the maximum force on both the grains and boundary vertices satisfies $\max_i(|{\vec f}_{i}|, |{\vec f}^b_i|) \leq 10^{-14}$~\cite{guenole_assessment_2020}. During energy minimization, the grain positions are allowed to change for all boundary conditions. The boundary vertices are allowed to change in response to forces arising from the grains in the FLEX boundary conditions, while ${\vec b}_i$ are held fixed for the FIX boundary conditions. We successively grow and shrink the grains until the pressure satisfies $\left| p- p_t \right| / p_t < 10^{-3}$, where $p_t$ is the target pressure~\cite{zhang_designing_2023, vanderwerf_pressure_2020}. Using this procedure, we generate $10^3$ sphere packings per system size $N$ at jamming onset, $p_t = 10^{-7}$.  
\subsection{Elastic modulus tensor}
\label{methods_b:c_tensor}

In $d$ dimensions, the elastic modulus tensor $C_{ijkl}$ is a 4th rank tensor that has at most $d(d+1)(d^2+d+2)/8$ independent elements. We will use Voigt notation, which expresses the 4th rank tensor as a 2nd rank tensor with $d(d+1)/2$ dimensions. $C_{mn}$ denotes the Voigt indices of the elastic modulus tensor, where both $n$ and $m$ range from $1$ to $6$ and correspond to the elementary strains. The map from the Cartesian indices $ij$ to the Voigt index $n$ is explicitly ($xx\mapsto 1$, $yy \mapsto2$, $zz \mapsto 3$, $yz \mapsto 4$, $xz \mapsto 5$, and $xy \mapsto 6$). In three dimensions, $C_{mn}$ is a $6\times6$ symmetric matrix. For disordered materials, which generally do not have discrete symmetries, we expect $21$ independent elastic constants. To obtain the full set of $21$ elastic moduli in 3D, we apply each of the six elementary strains $\gamma_{n}$, which correspond to the three compressive strains and three shear strains. (See Appendix \ref{si:strains}.) For a simple shear deformation in the $x$-$y$ plane, the strain tensor is 
\begin{equation}
\mathbf{\Gamma_6} = \begin{bmatrix}
0 & \gamma_{xy} & 0\\
0 & 0 & 0 \\
0 & 0 & 0
\end{bmatrix},
\end{equation}
where $\gamma_{xy}$ (in Voigt notation, this corresponds to $\gamma_6$) is the shear strain. During the strain test, we apply affine simple shear strain steps of $\gamma_{xy} = 10^{-9}$ to both the grains and boundary vertices, i.e. ${\vec r}_{\gamma i}={\vec r}_i \,(\mathbf{\Gamma+ \mathbf{I}})$ and ${\vec b}_{\gamma i}={\vec b}_i \,(\mathbf{\Gamma+ \mathbf{I}})$, where $\mathbf{I}$ is the identity matrix, ${\vec r}_{\gamma i}$ and ${\vec b}_{\gamma i}$ are the strained grain and boundary vertex positions. Following the application of each shear strain step, we minimize the total potential energy for the grain positions, keeping the boundary fixed, until $\max_i(|{\vec f}_{i}|, |{\vec f}^b_i|) \leq 10^{-14}$. We then calculate the stress $\Sigma_n$ (in Voigt notation) using the virial expression in Eq.~\ref{eqn: virial} after each strain increment, and obtain the elastic modulus by evaluating the slope $C_{mn} = d\Sigma_{n}/d \gamma_{m}$. 

We will calculate the full set of elastic moduli, $C_{mn}$, over a range of target pressures $p_t$, by sampling $30$ values logarithmically spaced between $10^{-7}$ and $10^{-2}$. Sampling pressures above jamming onset is similar to generating a static packing at jamming onset. We grow and shrink the grains with successively smaller packing fraction increments until the pressure satisfies $\left| p- p_t \right| / p_t < 10^{-3}$. We do not allow the boundary to deform during this process to ensure that the shape of the boundary remains similar to its shape at jamming onset. 
\begin{figure*}
\centering
\includegraphics[width = 2.0\columnwidth]{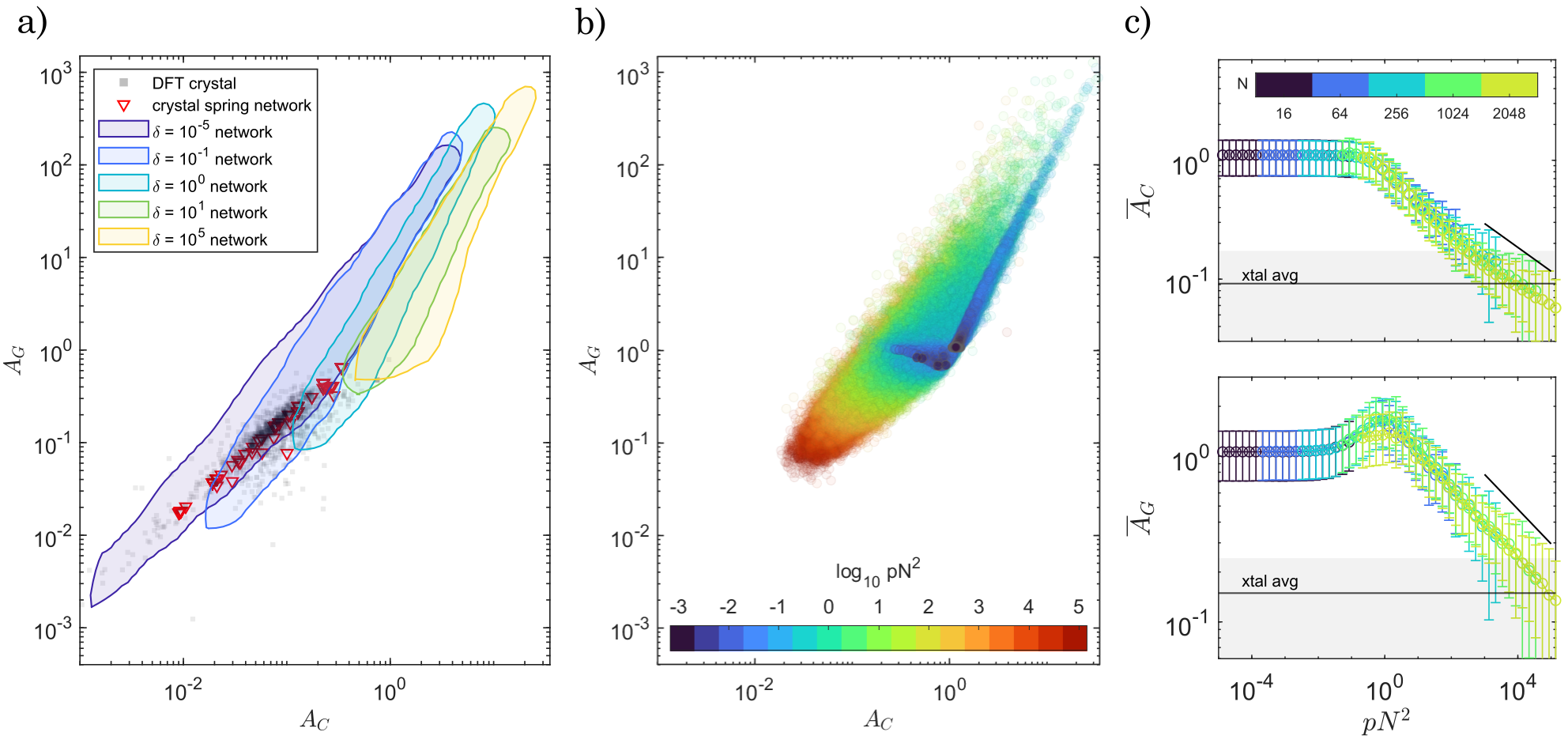}
\caption{(a) Shear anisotropy $A_G$ plotted versus compressive anisotropy $A_C$ for $1181$ atomic crystals (gray filled squares) and $448$ cubic spring networks derived from these crystalline materials (red triangles). We also show data from randomly generated elastic modulus tensors with different amounts of crystalline symmetry, ranging from strongly cubic (small $\delta$; purple shaded region) to strongly triclinic (high $\delta$; yellow shaded region). For randomly generated networks, the shaded regions correspond to a probability density of $p\geq10^{-4}$ smoothed with a Savitzky-Golay filter of order $2$ and window size $9$. The elastic moduli for the crystalline data are calculated from ab initio simulations\protect{~\cite{de_jong_charting_2015}}. (b) Shear anisotropy $A_G$ plotted versus compressive anisotropy $A_C$ for granular packings in periodic boundary conditions, colored by $pN^2$ increasing from violet to red.
(c) [Top] Ensemble-averaged compressive anisotropy ${\overline A}_C$ and [bottom] shear anisotropy ${\overline A}_G$ plotted as a function of $pN^2$ for granular packings in periodic boundary conditions. The ensemble average is taken over configurations at fixed $N$ and $p$. The data is colored from purple to yellow with increasing system size $N$, and the error bars indicate the standard deviation. The power-law tails in ${\overline A}_C$ and ${\overline A}_G$ at large $pN^2$ have slope $-0.20\pm 0.01$. Also shown are the average compressive and shear anisotropy (black lines) and one standard deviation from the average (grey region) for the crystal database.}
\label{fig:aniso_plane}
\end{figure*}

\begin{figure*}
\centering
\includegraphics[width = 2\columnwidth]{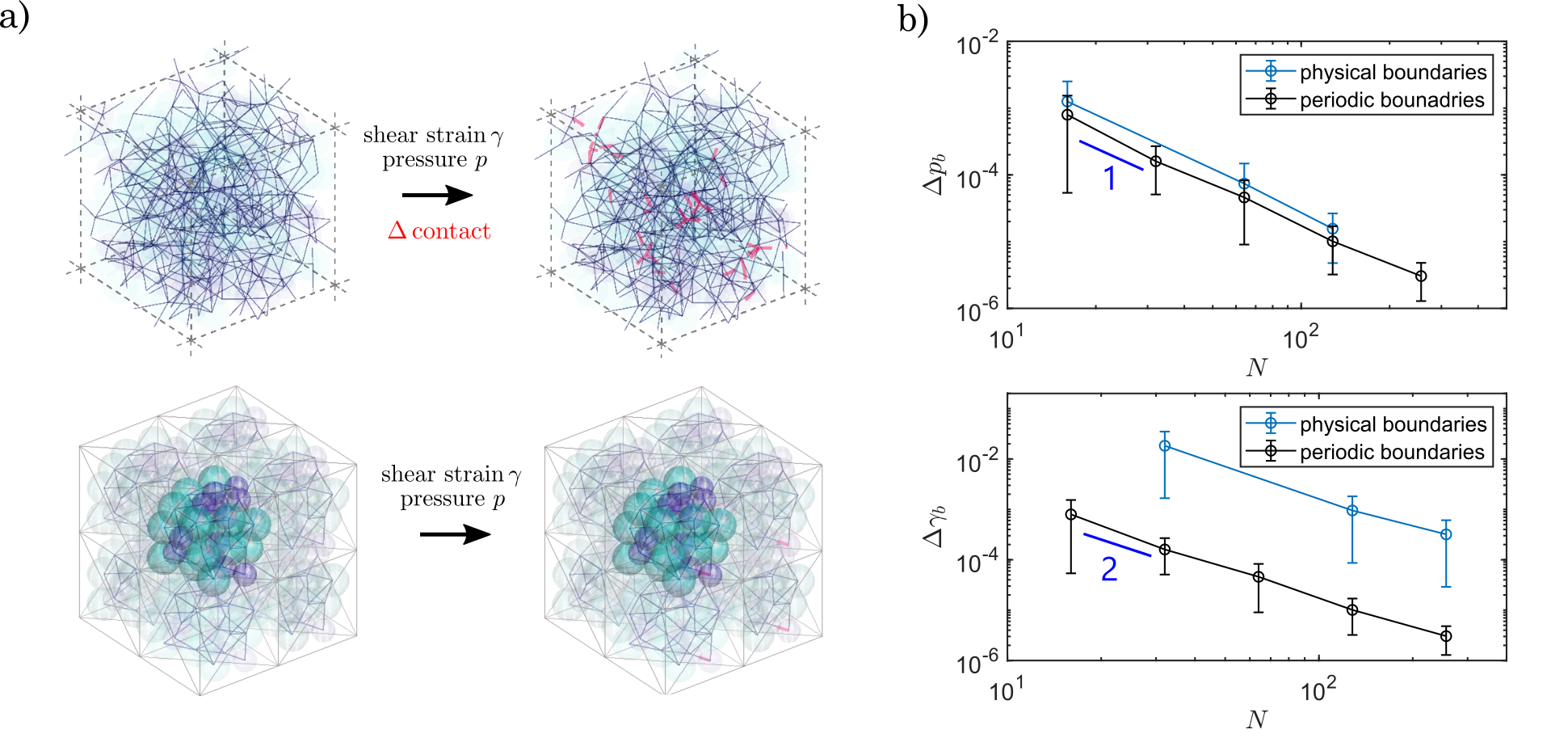}
\caption{(a) (Top) Granular packing with $N=256$ in periodic boundary conditions. The black lines show intergrain contacts. After changing the pressure by $\Delta p$ or applying a shear strain increment $\Delta \gamma$, the granular packing undergoes both contact breaks and additions, which are highlighted in red. (Bottom) Tessellated granular packing with $8$ single voxels of $N_{\textrm{i}}=32$ grains each, with $N=256$ total grains. Contact changes are highlighted in red.
(b) Pressure $\Delta p_{b}$ at which the first contact change occurs plotted as a function of the total number of grains $N$ for systems with (black line) periodic boundary conditions and (blue line) FIX cubic boundaries. (c) Shear strain $\Delta \gamma_{b}$ at which the first contact change occurs for the same systems in (b). In (b) and (c), the initial jammed packings have pressure $p_t=10^{-7}$.}
\label{fig:rearrangements}
\end{figure*}

\subsection{Measures of elastic anisotropy}
\label{methods_c:anisotropy}

For polycrystals, crystalline materials with large defect concentrations, and amorphous materials, the elastic response can be highly anisotropic. We calculate rotationally-invariant measures of elastic anisotropy using a similar approach to that in previous work~\cite{goodrich_jamming_2014, zhang_local_2023}. First, we define $R_{\Gamma}=\Delta U/V$ as the linear response of the system to an applied deformation $\Gamma$:
\begin{equation}
\label{eqn:R_gamma}
    R_\Gamma = C_{mn} \Gamma_m\Gamma_n.
\end{equation}
However, both $R_\Gamma$ and $C_{mn}$ depend on the coordinate system, so that
\begin{equation}
    R_\Gamma(\theta, \varphi, \psi) = C_{mn}(\theta, \varphi, \psi) \Gamma_m\Gamma_n,
\end{equation}
where $\theta$, $\varphi$, and $\psi$ are the Euler angles in the ZXZ' rotation sequence. To describe the transformation of $C_{mn}$ under changes of coordinate system, we will write out all $4$ indices in Cartesian coordinates, as $C_{ijkl}$. Converting between $C_{ijkl}$ and $C_{mn}$ can be carried out using the following mapping: ($xx$ corresponds to $1$, $yy$ to $2$, $zz$ to $3$, $yz$ to $4$, $xz$ to $5$, and $xy$ to $6$). $C_{ijkl}$ transforms under rotations of the coordinate system as
\begin{equation}
     C_{ijkl}(\theta, \varphi, \psi) =C_{pqrs}T_{ip}T_{jq}T_{kr}T_{ls},
\end{equation}
where $C_{pqrs}$ is the elastic modulus tensor in an arbitrary reference frame, $T$ is the $3\times3$ rotation matrix defined by $\theta$, $\varphi$, and $\psi$, and $C_{ijkl}$ is the elastic modulus tensor in the rotated frame. We define the mean shear modulus $G_{DC}$ and amplitude of the deviation from the average, $G_{AC}$. The angle-averaged shear modulus is 
\begin{equation}
G_{DC} = \frac{1}{8\pi^2} \int_0^{2\pi} \int_0^\pi \int_0^{2\pi}  R_{\Gamma_6}(\theta, \varphi, \psi) \, d\theta \, \sin\varphi \, d\varphi \, d\psi, 
\label{eqn: integral}
\end{equation}
where $R_{\Gamma_6}$ is the change in energy after an applied shear strain. Note that the integral of $R_{\Gamma_4}$ and $R_{\Gamma_5}$ are the same as that in Eq.~\ref{eqn: integral}, so that the choice of $R_{\Gamma_6}$ is not arbitrary. We also calculate the angle-averaged variance in the shear modulus:
\begin{widetext}
\begin{equation}   
\label{eqn: integral_ac}
G_{AC}^2 = \frac{1}{8\pi^2} \int_0^{2\pi} \int_0^\pi \int_0^{2\pi}  [R_{\Gamma_6}(\theta, \varphi, \psi) - G_{DC}]^2 \\
\, d\theta \,\sin\varphi\, d\varphi \, d\psi. 
\end{equation}
\end{widetext}
We define the normalized shear anisotropy as the ratio of the standard deviation over the average value, $A_G=G_{AC}/G_{DC}$. We define ${\overline A}_G=\langle G_{AC} \rangle / \langle G_{DC} \rangle$ as the ensemble-averaged shear anisotropy. Similarly, we define the compressive anisotropy by integrating $R_{\Gamma_C}$ for the compressive strain $\Gamma_C=(\Gamma_1+\Gamma_2+\Gamma_3)/3$ over all angles. To calculate the angle-averaged bulk modulus $B_{DC}$, we integrate $R_{\Gamma_U}$ for a uniaxial compressive strain $\Gamma_U=\Gamma_1$. We also determine the uniaxial compressive variance, $U_{AC}$ using $R_{\Gamma_U}$ and $U_{DC}$ in Eq. \ref{eqn: integral_ac}. From these quantities, we define the normalized compressive anisotropy $A_C=U_{AC}/(G_{DC}B_{DC})^{1/2}$ and the ensemble-averaged compressive anisotropy $\overline{A}_C = \langle U_{AC} \rangle /\langle (B_{DC} G_{DC})^{1/2}\rangle$. Note that a perfectly isotropic material has $A_G=A_C=0$.

Analytical expressions for $A_G$ and $A_C$ for an arbitrary triclinic material in terms of elements of the elastic modulus tensor $C_{mn}$ are provided in Appendix \ref{si:analytics}. Orientation averages can be evaluated on the elastic modulus tensor $C_{mn}$ (Voigt average) or the compliance tensor $S_{mn}=(C^{-1})_{mn}$ (Reuss average), corresponding to a strain-controlled and stress-controlled description respectively~\cite{hendrix_self-consistent_1998, man_simple_2011}. Another coordinate-invariant measure of anisotropy is the universal anisotropy index $A^U$, which can be written in terms of the ratios $G_{DC}({\bf C})/G_{DC}({\bf S})$ and $U_{DC}({\bf C})/U_{DC}({\bf S})$~\cite{ranganathan_universal_2008, kube_elastic_2016}. In Appendix \ref{si:au}, we compare results for $A_G$, $A_C$, and $A^U$.

\section{Results}
\label{sec:results}

In this section, we will characterize the shear and compressive anisotropy of atomic crystals, granular packings, and tessellated granular materials. In Sec.~\ref{results_a:pbc}, we show that atomic crystals have relatively small elastic anisotropy, while granular packings can achieve values up to $10^2$ times larger when $pN^2\sim1$. However, elastic anisotropy homogenizes when grain contacts change in response to an applied strain. In Sec.~\ref{results_b:tessellation}, we show that single voxels, consisting of a small number of grains within physical boundaries, prevent grain rearrangements and possess large $A_G$ and $A_C$. We further show that homogeneous tessellations, where each voxel has the same grain configuration, can inherit the single voxel response with large $A_G$ and $A_C$. We also demonstrate that heterogeneous tessellations with two different single voxel grain configurations and homogeneous tessellations with extra boundary degrees of freedom can also attain large $A_C$ and $A_G$

\subsection{Elastic anisotropy of individual granular packings}
\label{results_a:pbc}

We begin by calculating the shear and compressive anisotropy for atomic crystalline materials from the large database in Ref.~\cite{de_jong_charting_2015}. This database contains $1181$ crystalline inorganic compounds, including metals, metallic compounds, semiconductors, and insulators. In order of decreasing symmetry, the database consists of $452$ cubic, $239$ hexagonal, $59$ trigonal, $193$ tetragonal, $193$ orthorhombic, $45$ monoclinic, and $0$ triclinic crystals. The database includes the full elastic modulus tensor $C_{mn}$ for these materials obtained from first-principles calculations using density functional theory (DFT) which are validated against experimental measurements. In addition, we have generated spring networks in (cubic) periodic boundary conditions derived from the $452$ cubic crystal structures in the database. Using the atom coordinates, atom species, and lattice vectors provided from the database, we create a $2\times2\times2$ supercell of the unit cell and determine bonds between atoms with the CrystalNN method~\cite{pan_benchmarking_2021}. The initial positions of the bonded atoms determine the rest length of each spring. We perform successive strain tests to the spring networks as detailed in Sec.~\ref{methods_b:c_tensor} to calculate the elastic modulus tensor. (In Appendix~\ref{si:spring_nets}, we describe the cubic crystal derived spring networks in more detail.) In Fig. \ref{fig:aniso_plane} (a), we plot the shear anisotropy $A_G$ versus the compressive anisotropy $A_C$ for the atomic crystalline materials, spring networks derived from the crystalline materials, and randomly generated elastic modulus tensors. (In Appendix~\ref{si:stability}, we provide additional details concerning the randomly generated elastic modulus tensors.) We find that both the crystalline materials and the spring networks derived from the crystalline materials possess small values for $A_C$ and $A_G$; on average, $\overline{A}_G \approx 0.15$ and $\overline{A}_C \approx 0.09$. This result emphasizes that the small values for the anisotropy of crystalline materials is independent of the specific atomic interactions and that structural order is important for determining the anisotropy of solids. To probe the role of structural symmetry, we control the relative strength of the symmetric and asymmetric contributions to the elastic modulus tensor. We generate $10^6$ random elastic tensors $C^{\textrm{rand}}_{mn} = C^{\textrm{cubic}}_{mn} + C^{\textrm{triclinic}}_{mn}$ composed of a symmetric cubic component ($3$ independent constants) and an asymmetric triclinic component ($21$ independent constants) whose relative strength is tuned by $\delta=C^{\textrm{cubic}}_{mn}/C^{\textrm{triclinic}}_{mn}$. We normalize ${\bf C}$ so that its $L_2$ norm is unity, i.e. $\|{\bf C}\|_2= \max_{i} \sqrt{c_i {\bf C}^T {\bf C}}$ where $c_i$ are the eigenvalues of ${\bf C}$. (See Appendix~\ref{si:stability} for a discussion of the stability of randomly generated ${\bf C}$.) We show the probability distribution $P(A_C,A_G) \ge P_0$ for randomly generated elastic modulus tensors in Fig.~\ref{fig:aniso_plane} (a)  (where $P_0 =10^{-4}$) over a range of $\delta$ from $\delta=10^{-5}$ (purple) to $10^{5}$ (yellow). Although increasing symmetry tends to lower both the compressive and shear anisotropy, it is possible to obtain large values of anisotropy in a strongly cubic systems. This result implies that there is an interplay between a material's structural symmetry, which controls the number of independent elements in the elastic modulus tensor, and the magnitudes of the independent elastic elements, that together determine the elastic anisotropy. 

Next, we examine $A_G$ and $A_C$ for jammed granular packings in periodic boundary conditions, which have no structural symmetry and are triclinic. In Fig.~\ref{fig:aniso_plane} (b), we plot $A_G$ and $A_C$ for $10^3$ individual packings over a wide range $pN^2$ with $16 \le N \le 2048$ and $30$ values of pressure in the range $10^{-7} \le p \le 10^{-2}$. Granular packings with lower $pN^2$ have larger elastic anisotropies up to $A_C\sim 10^2$ and $A_G\sim 10^3$. For $pN^2\gtrsim 10^3$, the elastic anisotropy decreases and approaches the region spanned by the crystal database $A_C \sim A_G\sim10^{-1}$. In general, for granular packings, increases and decreases in either $A_G$ or $A_C$ correlate with similar changes in the other measure of elastic anisotropy. Even though granular packings are considered triclinic in the sense that they have no discrete symmetries, $A_C$ and $A_G$ for granular packings with $pN^2\geq10^2$ are smaller than those for randomly generated elastic modulus tensors with $\delta=10^5$ in Fig.~\ref{fig:aniso_plane} (a). This result shows that although increasing $pN^2$ preserves the structural symmetry of a packing, it increases the relative strength of the symmetric elastic modulus contribution. 

To emphasize the $pN^2$ dependence of the elastic anisotropy for granular packings, in Fig.~\ref{fig:aniso_plane} (c), we plot the ensemble averages $\overline{A}_G$ and $\overline{A}_C$ versus $pN^2$. We also indicate the average values $\overline{A}_G$ and $\overline{A}_C$ plus or minus one standard deviation for atomic crystalline materials from the crystal database, which do not vary with $pN^2$. For granular packings, both $\overline{A}_G$ and $\overline{A}_C$ collapse with $pN^2$ and exhibit a plateau ${\overline A}_C \sim {\overline A}_G \sim 1$ at small $pN^2\leq10^{-2}$. In this regime, the granular packings are isostatic. In the large $pN^2$ limit, $\overline{A}_G$ and $\overline{A}_C$ decrease toward ${\overline A}_C \sim {\overline A}_G \sim 0$ with a power-law scaling exponent $-0.20\pm0.02$, similar to previous work~\cite{goodrich_jamming_2014, zhang_local_2023}. As the pressure $p$ is increased at fixed $N$, granular packings undergo changes in their contact networks and homogenize, eventually reaching small elastic anisotropies that are characteristic of crystalline solids. In particular, for $pN^2\geq10^4$, granular packings have similar average elastic anisotropy as crystalline materials. To create materials with strong elastic anisotropy, we will tessellate small granular packings with $pN^2\sim1$, which typically have one extra contact above isostaticity and possess the largest values of $A_C$ and $A_G$~\cite{zhang_local_2023}. In this regime, typical granular packings have elastic anisotropy that are at least $10$ times more anisotropic than a typical crystalline material. 

Although granular packings are strongly anisotropic at $pN^2\sim1$, they rearrange under applied shear strain and compression, which causes homogenization in their elastic response~\cite{goodrich_jamming_2014}. To prevent grain rearrangements, we will build large anisotropic granular packings by tessellating many single small voxels with large elastic anisotropy, as shown in Fig.~\ref{fig:rearrangements} (a). To quantify the system size dependence of grain rearrangements, we start with a granular packing at jamming onset $p_t=10^{-7}$, and increase pressure by $10^2$ logarithmically spaced increments of $\Delta p$ from $10^{-8}$ to $10^{-1}$. We determine the smallest compressive perturbation $\Delta p_{b}$ that causes a contact change. In Fig.~\ref{fig:rearrangements} (b), we plot $\Delta p_{b}$ versus system size $N$ and show that it scales as a power-law with exponent $-2$. Similarly, we calculate the smallest shear strain $\Delta \gamma_{b}$ that causes a contact change and find that $\Delta \gamma_{b} \sim N^{-1}$. Granular packings with periodic and physical boundaries have the same power-law scaling for $\Delta p_b$ and $\Delta \gamma_b$ with $N$. Tessellating small granular subsystems within physical boundaries will limit rearrangements, which has been observed in both simulations and experiments in two dimensions ~\cite{pashine_tessellated_2023, zhang_designing_2023}.

\begin{figure}
\centering
\includegraphics[width = 1\columnwidth]{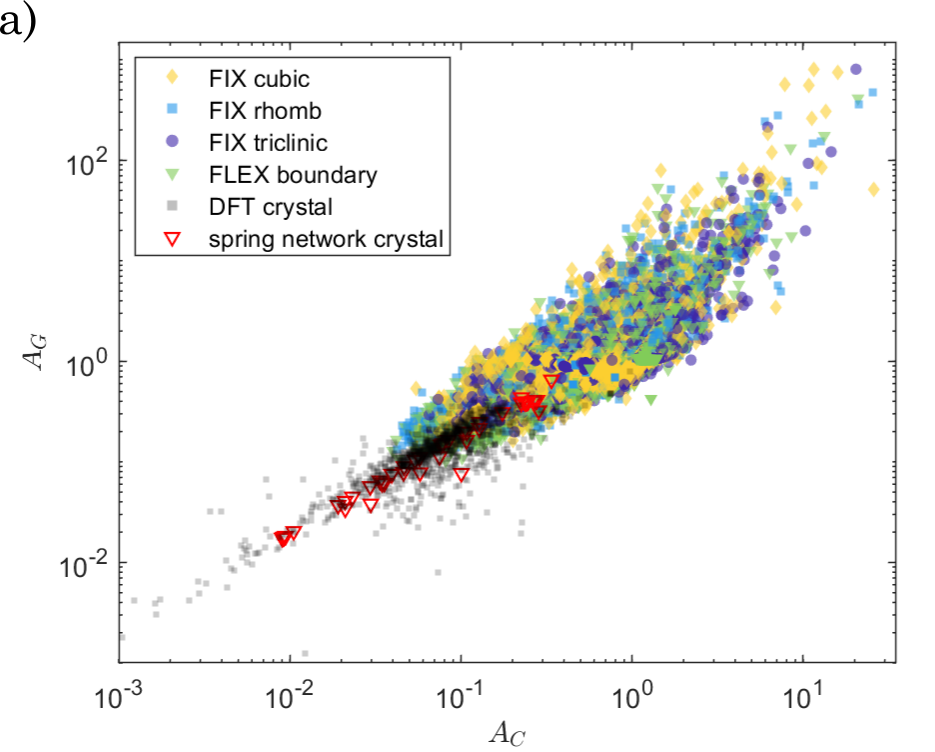}
\caption{Shear anisotropy $A_G$ plotted versus compressive anisotropy $A_C$ for granular packings over the range $10^{-6} \le pN^2 \le 10^4$ with FIX cubic (purple circles), FIX rhombohedral (blue squares), FIX triclinic (green triangles), and FLEX (yellow diamonds) boundary conditions. We also show data for crystalline materials from DFT calculations (gray filled squares) and spring networks derived from the crystalline materials (red triangles).}
\label{fig:single_vox}
\end{figure}

Next, we will verify that jammed granular packings confined within physical boundaries can attain large $A_C$ and $A_G$. In Fig.~\ref{fig:single_vox}, we show the shear and compressive anisotropy of granular packings  within physical boundary conditions (fixed triclinic, fixed rhombohedral, fixed cubic, and flexible) plotted versus $pN^2$.  Similar to the results for granular packings in periodic boundaries in Fig.~\ref{fig:aniso_plane} (b), granular packings within the four physical boundary conditions attain maximum $A_C\sim10^{1}$ and $A_G\sim10^2$ and minimum $A_C\sim10^{-1}$ and $A_G\sim10^{-1}$ that overlap with the values for atomic crystalline materials. Thus, we find that the range of $A_C$ and $A_G$ is not significantly changed by either the shape of the boundary for FIX boundary conditions or the ability to deform for FLEX boundary conditions.

\subsection{Elastic anisotropy of tessellations of granular packings}
\label{results_b:tessellation}

\begin{figure*}
\centering
\includegraphics[width = 2.0\columnwidth]{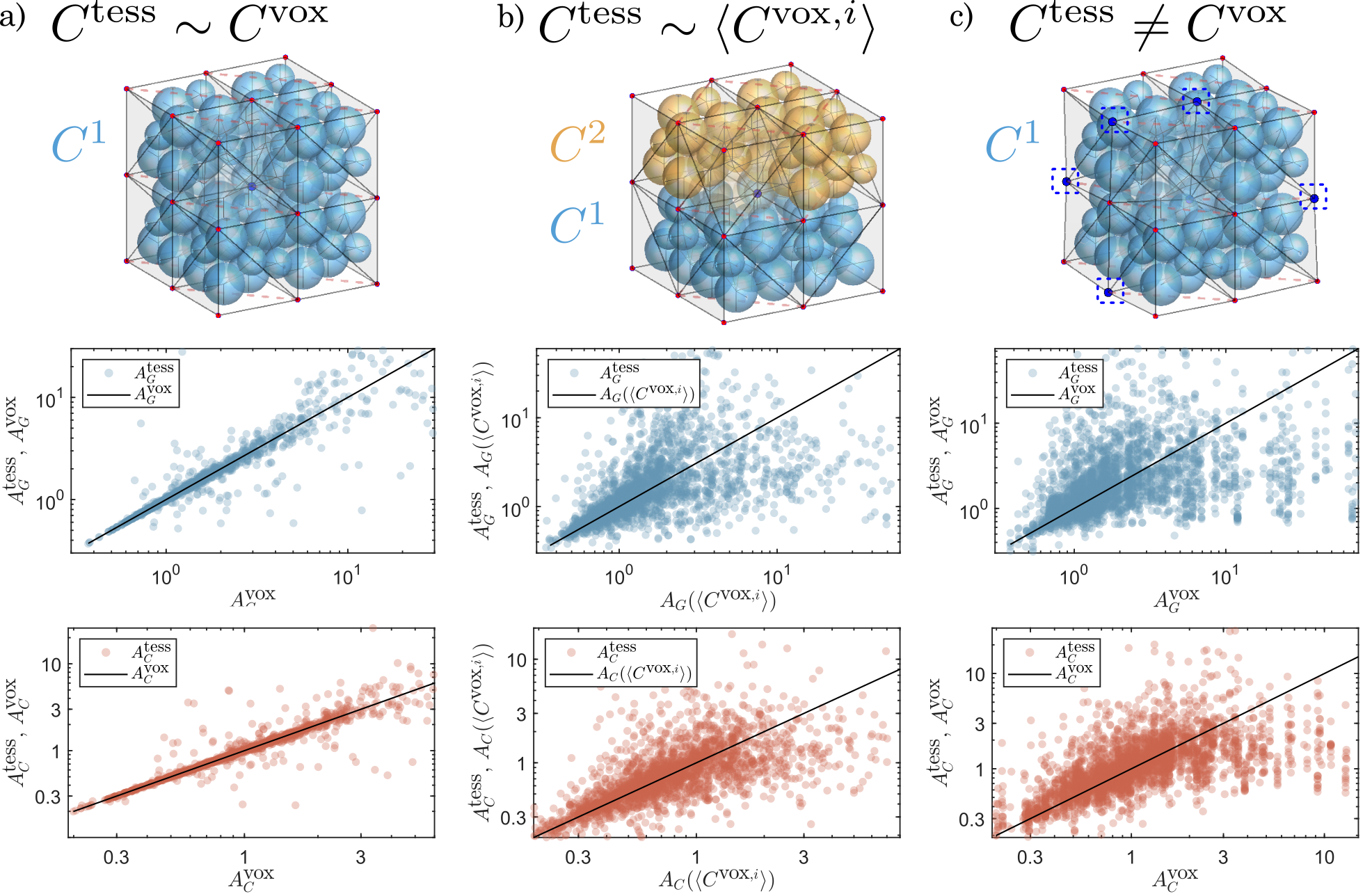}
\caption{(a) (Top) $2\times 2 \times 2$ homogeneous tessellations possess the same randomly generated $N_i=12$ granular packing with elastic modulus tensor ${\bf C}^1$ in each voxel with fixed surface vertices (red), while the single internal vertex is allowed to move in response to applied strain (blue). (Middle and bottom) Shear and compressive anisotropy of the homogeneous tessellations $A^{\textrm{tess}}_G$ and $A^{\textrm{tess}}_C$ plotted versus the individual voxel anisotropy $A^{\textrm{vox}}_G$ and $A^{\textrm{vox}}_C$. Each data point corresponds to a different granular packing in each single voxel of the homogeneous tessellation. The black dotted line indicates $A^{\textrm{tess}}_G=A^{\textrm{vox}}_G$ and $A^{\textrm{tess}}_C=A^{\textrm{vox}}_C$. (b) $2 \times 2 \times 2$ heterogeneous tessellations include two different granular packings with elastic modulus tensors ${\bf C}^1$ and ${\bf C}^2$. (Middle and bottom) Shear and compressive anisotropy of the heterogeneous tessellation $A^{\textrm{tess}}_G$ and $A^{\textrm{tess}}_C$ plotted versus the average shear and compressive anisotropy of the two granular packings in the single voxels, $A_G(\langle C^{\textrm{vox},i}\rangle)$ and $A_C(\langle C^{\textrm{vox},i}\rangle$). The black dotted line indicates  $A^{\textrm{tess}}_G=A_G(\langle C^{\textrm{vox},i}\rangle)$ and $A^{\textrm{tess}}_C=A_C(\langle C^{\textrm{vox},i}\rangle)$. 
(c) $2\times 2 \times 2$ homogeneous tessellation with $6$ randomly selected boundary vertices that can move in response to applied strain (blue square surrounding blue vertex). (Middle and bottom) Shear and compressive anisotropy of the tessellation $A^{\textrm{tess}}_G$ and $A^{\textrm{tess}}_C$ plotted versus the anisotropy of the voxels $A^{\textrm{vox}}_G$ and $A^{\textrm{vox}}_C$. The black dotted line indicates $A^{\textrm{tess}}_G=A^{\textrm{vox}}_G$ and $A^{\textrm{tess}}_C=A^{\textrm{vox}}_C$.
}
\label{fig:tess_aniso}
\end{figure*}

Since the boundary shape does not affect the range of elastic anisotropies that can be attained, we will focus on tessellations of cubic cells with FIX boundary conditions. We present results for three different types of tessellations. First, we show that homogeneous tessellations with the same granular packing in each voxel can inherit the elastic response of the single voxel. Next, we show that heterogeneous tessellations where the voxels include two different granular packings possess an elastic anisotropy that can fluctuate significantly around the average response of the two individual granular packings. Finally, we study homogeneous tessellations with external boundary degrees of freedom that are either constrained or unconstrained, which allows deformations of the tessellation boundary to couple with deformations of the grain packing to increase or decrease elastic anisotropy relative to the individual packing.

We study tessellations composed of single cubic voxels with $N_i=12$ grains. Two neighboring voxels share the square face between them and thus share $4$ vertices, $5$ edges, and $2$ triangular faces. As an example, $2\times2\times2$ tessellations include $N_v=27$ total vertices, $N_e=90$ edges (with $12$ removed length constraints diagonally in the $x$-$y$ plane), and $N_{tri}=72$ triangular faces.  To enforce the external structure of the cubic boundaries and lock in the mechanical response of the interior granular packings, we will first fix the boundary vertices on the exterior surface and allow only the internal boundary vertices to deform~\cite{zhang_designing_2023}. $2\times2\times2$ tessellations possess one internal vertex, while the remaining vertices are on the exterior surface, as shown in  Fig.~\ref{fig:tess_aniso} (a), where the fixed external vertices are colored red and the interior flexible vertex is colored blue. As another example, $3\times3\times3$ tessellations possess $N_v=64$ total vertices, $N_e=252$ edges (with $36$ removed length constraints diagonally in the $x$-$y$ plane), and $N_{tri}=216$ triangular faces. In $3 \times 3 \times 3$ tesselations, there are $8$ internal flexible vertices. In this work, we focus on $2\times2\times2$ tessellations, which is the smallest nontrivial tessellation in 3D. To maximize anisotropy, we tessellate single voxels with $pN^2 = 10^{-0.25}$, $1$, and $10^{0.25}$. In Fig.~\ref{fig:tess_aniso} (a), we show the elastic anisotropy of homogeneous tessellations constructed from $500$ different randomly generated jammed granular packings placed in each individual voxel for the three values of $pN^2$. We plot the compressive and shear anisotropy $A^{\textrm{tess}}_C$ and $A^{\textrm{tess}}_G$ for the homogeneous tessellations versus the single voxel anisotropy $A^{\textrm{vox}}_C$ and $A^{\textrm{vox}}_G$, where each point represents a particular single voxel and tessellation pair. We also indicate the lines where $A^{\textrm{tess}}_C=A^{\textrm{vox}}_C$ and $A^{\textrm{tess}}_G=A^{\textrm{vox}}_G$. We find that homogeneous tessellations possess shear and compressive anisotropies that are similar to the corresponding single voxel values. Deviations in the anisotropies from the single voxel values occur due to deformation of the single internal vertex in response to applied strain. Homogeneous tessellations are able to harness the anisotropy of granular packings in single voxels to yield large systems with large elastic anisotropy. 

In addition to homogeneous tessellations, we also investigate heterogeneous tessellations composed of two different granular packings. The heterogeneous tessellations are constructed by placing one type of granular packing with elastic modulus tensor ${\bf C}^1$ in the bottom layer and a different packing with ${\bf C}^2$ in the upper layer as shown in Fig.~\ref{fig:tess_aniso} (b). $2000$ pairs of granular packings are randomly sampled without replacement from the original $500$ packings discussed in Fig.~\ref{fig:tess_aniso} (a). We expect that the elastic modulus tensor of the heterogeneous tessellation will be close to an element-wise average of the two elastic modulus tensors of the two constituent packings. Note that $A_G$ and $A_C$ are nonlinear functions of the elements $C_{mn}$ so that $A^{\rm tess}_G$ and $A^{\rm tess}_C$ can be significantly larger or smaller than $A^{\rm vox}_G$ and $A^{\rm vox}_C$ of the constituent packings. In Fig.~\ref{fig:tess_aniso} (b), we plot $A^{\textrm{tess}}_C$ and $A^{\textrm{tess}}_G$ versus the average anisotropy of the two constituent single voxels, $A_C(\langle {\bf C}^{\textrm{vox},i}\rangle)=A_C(({\bf C}^{\textrm{vox},1}+{\bf C}^{\textrm{vox},2})/2)$ and $A_G(\langle {\bf C}^{\textrm{vox},i}\rangle)=A_G(({\bf C}^{\textrm{vox},1}+{\bf C}^{\textrm{vox},2})/2)$.  The heterogeneous tessellations do, on average, have shear and compressive anisotropy that are similar to the single voxel averages, with fluctuations from the motion of the interior vertex. However, heterogeneous tessellations offer more tunability than homogeneous tesselations. For example, we can select ${\bf C}^{\textrm{vox},1}$ and ${\bf C}^{\textrm{vox,2}}$ for packings in the single voxels so that the nonlinear functions $A_C$ and $A_G$ of $({\bf C}^{\textrm{vox},1} + {\bf C}^{\textrm{vox},2})/2$ are larger than the individual voxel values. This is a way to effectively enhance the anisotropy of the tessellation by way of configurational heterogeneity.

Finally, we explore the effect of removing some of the external boundary constraints and allowing boundary deformations in response to applied strain. We randomly select $6$ exterior vertices to become flexible for $10$ random seeds for each of the $500$ tessellations. We do not report tessellations that become unjammed. In Fig.~\ref{fig:tess_aniso} (c), we plot the tessellation anisotropies $A^{\textrm{tess}}_G$ and $A^{\textrm{tess}}_C$ versus the single voxel anisotropies $A^{\textrm{vox}}_G$ and $A^{\textrm{vox}}_C$ for the flexible homogeneous tessellations. In the range $A_U^{\textrm{vox}}\leq3$ and $A_G^{\textrm{vox}}\leq10$, the elastic anisotropy for the flexible tessellations scale with those for the single voxels, with fluctuations due to internal and external boundary deformations. The flexible tessellations provide another source of tunability for the elastic anisotropy since removing the constraints of certain boundary vertices of the tessellations can enhance the anisotropy of the single voxels. At larger values of $A^{\textrm{vox}}_C$ and $A^{\textrm{vox}}_G$, the boundary deformations in flexible tessellations homogenize the elastic anisotropy in the single voxels. Even with this average homogenization, we find that particular tessellations can achieve large values $A^{\rm tess}_G\sim10^2$ and $A^{\rm tess}_C\sim10$. Note that all three tessellation modalities (homogeneous with boundary constraints, heterogeneous, and homogeneous with flexible boundaries) can attain $A_G$ and $A_C$ that are significantly larger than the values for crystalline solids. 

\section{Conclusions and future directions}
\label{sec:conclusions}

In this article, we show that making tessellations of small granular packings, effectively making a cubic crystal with a disordered basis, yields bulk granular materials that have large elastic anisotropy and limits grain rearrangements that homogenize the elastic response in conventional granular solids. This work is a step forward in realizing directionally targeted elastic properties, which goes beyond metamaterial designs that have largely focused on tuning isotropic elastic properties. We first studied the elastic anisotropy of individual granular packings. We quantified elastic anisotropy using two normalized and rotational-invariant quantities that capture shear anisotropy, $A_G=G_{AC}/G_{DC}$ and compressive anisotropy $A_C=U_{AC}/(G_{DC}B_{DC})^{1/2}$. In particular, granular packings in systems with periodic boundary conditions at intermediate values of $pN^2\sim1$ possess large values of $A_G\sim{10^2}$ and $A_C\sim{10^1}$ around two orders of magnitude larger than that for crystalline solids. In the large $pN^2$ limit, granular packings in periodic boundary conditions become isotropic and have comparable anisotropy to that of crystalline solids. 

Next, we discuss the properties of granular packings with physical boundaries, i.e. the voxels or subunits that comprise the tessellated materials. We study three different fixed boundary shapes, as well as a flexible boundary with a single zero-energy shear mode. For the $pN^2$ values considered, the physical boundary conditions do not significantly impact the elastic anisotropy of the granular packings and the results are similar to those for periodic boundary conditions. In particular, at $pN^2\sim1$, granular packings with physical boundaries possess large $A_G\sim10^2$ and $A_C\sim10^1$. 

We considered three types of tessellations of small granular packings: homogeneous, heterogeneous, and homogeneous with flexible boundaries. All of these tessellations limit grain rearrangements in each voxel. For the simplest case of a single cubic voxel repeated multiple times in a homogeneous tessellation with boundary constraints, the elastic anisotropy for the tessellation is similar to that for a single voxel. For heterogeneous tessellations, where the single voxels contain one of two possible granular packings, the tessellation posesses elastic anisotropy that is roughly determined by ${\bf C}^{\textrm{tess}}\approx ({\bf C}^1 + {\bf C}^{2})/2$, where ${\bf C}^1$ and ${\bf C}^2$ are the elastic modulus tensors of the individual grain packings. However, since $A_G$ and $A_C$ are nonlinear functions of the individual elastic modulus tensor elements, $C_{mn}$, $A_G$ and $A_C$ can be larger than the single voxel values. Finally, for a homogeneous tessellation with flexible boundary vertices, we find that boundary deformations can increase (or decrease) the elastic anisotropy compared to that for single voxels. In general, tessellations of granualr packings can preserve the large elastic anisotropy of the constituent granular packings. However, in many cases, we can enhance the elastic anisotropy of individual voxels by adding a heterogeneous collection of grain packings and boundary flexibility. For all three tessellation modalities, we obtain large elastic anisotropies up to $A_G\sim10^2$ and $A_C\sim10^1$, which are two orders of magnitude larger than the elastic anisotropies for crystalline materials.

This work raises several interesting directions for future research. Here, we focused on the simplest heterogeneous tessellations where half of the voxels contain one grain packing and the other half contain the other. However, even for small tessellations containing $8$ voxels, the number of distinct heterogeneous tessellations for $N_i=12$ grain packings is large. Previously, genetic algorithms have been implemented to efficiently explore the large space of configurations and find novel material designs in the context of granular metamaterials, solid state materials, and mechanical metamaterials~\cite{parsa_universal_2023, le_discovery_2016, zunger_inverse_2018}. This approach will be very useful in the context of designing heterogeneous tessellations, allowing optimization of desired elastic modulus tensor $C_{mn}$. These methods can also explore doubly heterogeneous tessellations, i.e. tessellations that contain voxels with different shapes and sizes, where each voxel contains different jammed granular packings. 

While experimental realizations of tessellated granular materials have been realized in 2D, many challenges exist for creating 3D tessellations in experiments. First, voxels need to have a boundary that is strong enough to prevent the granular packing from rearranging, but also flexible enough so that the stiffness of the boundary does not dominate the grains contribution to the total elastic response. Moreover, our results for flexible tessellations in Fig.~\ref{fig:tess_aniso} (c) suggest that its elastic anisotropy can vary sensitively with the specific granular packing and physical boundaries. An experimental 3D voxel is also required be a shape that can fill space, and thus out of plane boundary buckling modes must be suppressed or strongly coupled to neighboring voxels. Also, experimentally measuring all elements of the elastic modulus tensor is challenging. Existing methods in materials science and the geophysical sciences have used acoustic scattering experiments to measure the compression and shear wave velocities over a range of orientations~\cite{donahue_measuring_2019, fan_determination_2015}. However, these methods can be sensitive to sample geometry and require large signal to noise ratios when applied to low symmetry materials such as granular packings. Another interesting extension of this work is to find coordinate invariant measures of anisotropy that can be calculated from experimentally tractable measurements of the Young's modulus and shear modulus, rather than difficult to measure moduli, such as $C_{36}$ or $C_{16}$. This research thread aligns with existing open questions about determining the ``closest'' coordinate-invariant symmetric approximations for asymmetric tensors.

\section{Acknowledgments}

The authors acknowledge support from NSF Grant No. DMREF 2118988. This work was also supported by the High Performance Computing facilities operated by Yale's Center for Research Computing. 

\appendix

\section{Expressing the FLEX boundary potential in terms of boundary vertex positions}
\label{si:box_potential}

In Eq. \ref{eqn:uw} in the main text, we describe the shape-energy potential for the FLEX boundary condition as a function of the edge lengths $l_j$, triangle face areas $a_i$, volume $v$, and bending angles $\theta_k$ of the voxel boundary. In this Appendix, we explicitly write the edge length, triangle face area, volume, and bending energies as a function of the positions of the boundary vertices ${\vec b}_i$. For an edge connecting boundary vertex $i$ and $j$ the corresponding length is $l_{ij}=|b_i-b_j|^2$. For a triangular area element along boundary vertices $i$, $j$, and $k$, the area is $a = |({\vec b}_i-{\vec b}_j) \times ({\vec b}_j-{\vec b}_k)|$. The volume of a voxel is calculated using the shoelace formula by summing over all triangular faces $i$, $j$, and $k$ oriented such that the normal vectors are pointing outward: $v=\frac{1}{6}\sum_{i=1}^{N_f}\det \{{\vec b}_i,{\vec b}_j,{\vec b}_k\}$, where $N_f$ is the number of triangular faces and $\{{\vec b}_i,{\vec b}_j,{\vec b}_k\}$ is a $3\times3$ matrix whose columns are the vertex position vectors ${\vec b}_i, {\vec b}_j$, and ${\vec b}_k$. For two adjacent triangles forming a kite $ijkl$, where the shared edge is $jk$ and the individual triangles are $ijk$ and $jkl$, the bending angle between them is $\theta_{ijkl}=\arcsin (\hat{n}_{ijk}-\hat{n}_{jkl}) \, \textrm{sgn}(\theta_{ijkl})$, $\hat{n}_{ijk} = (({\vec b}_i-{\vec b}_j)\times({\vec b}_j-{\vec b}_k))/|({\vec b}_i-{\vec b}_j)\times({\vec b}_j-{\vec b}_k)|$ is the unit normal vector of triangle $ijk$ and $\hat{n}_{jkl} = (({\vec b}_j-{\vec b}_k)\times({\vec b}_k-{\vec b}_l))/ |({\vec b}_j-{\vec b}_k)\times({\vec b}_k-{\vec b}_j)|$ is the unit normal vector of triangle $jkl$. The sign of the $\theta_{ijkl}$ is defined to be $\textrm{sgn}(\theta_{ijkl})= \textrm{sgn}((\hat{n}_{ijk}\times\hat{n}_{jkl})\cdot(\vec{b_i}-\vec{b_j}))$ which is negative when the edge $kl$ is bent inward and positive when the edge $kl$ is bent outward.

\begin{figure}
\centering
\includegraphics[width = 1.0\columnwidth]{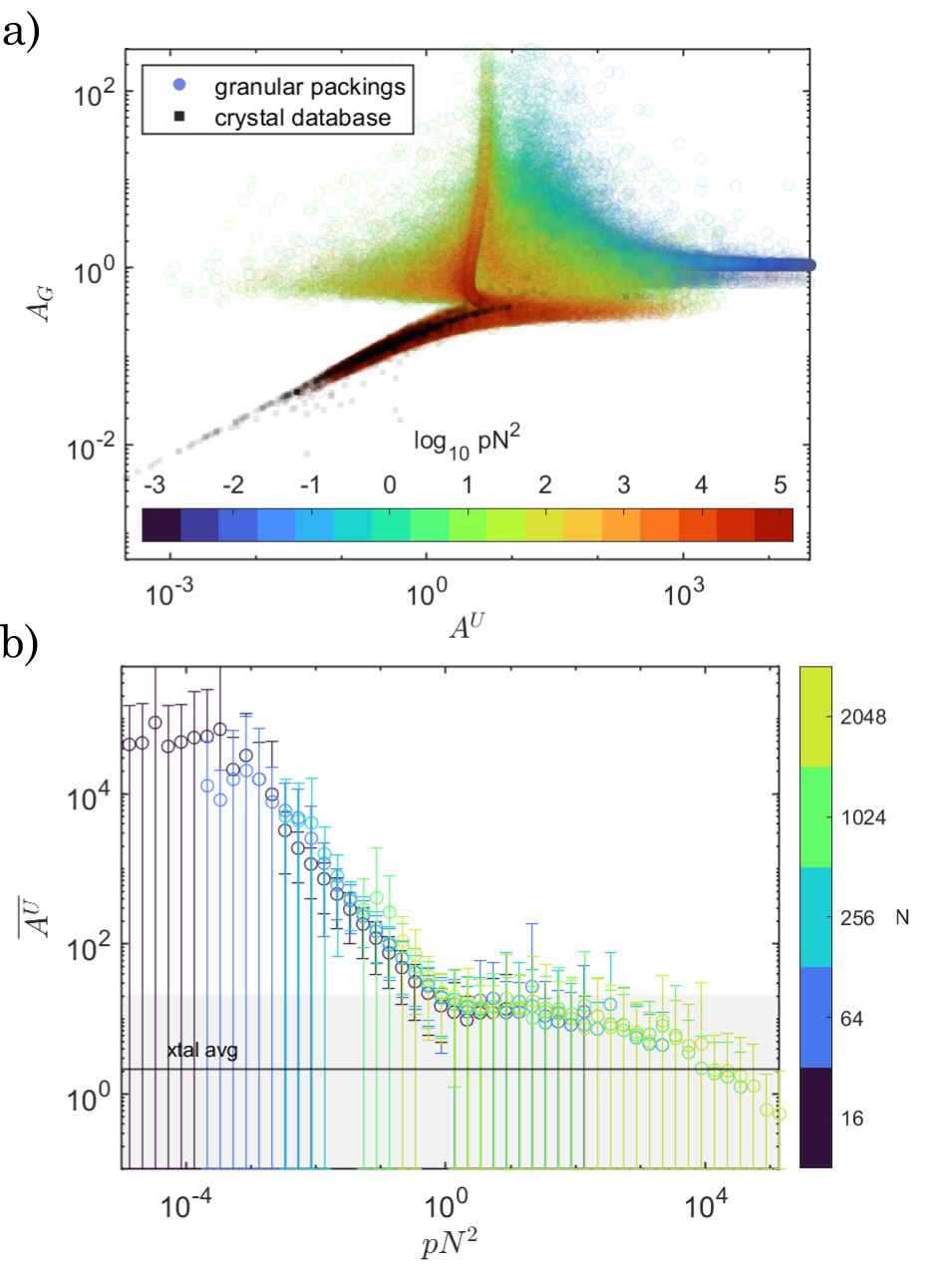}[!htb]
\caption{
(a) The shear anisotropy $A_G$ plotted versus the universal anisotropy index $A^U$ for granular packings in periodic boundary conditions (circles) colored by $pN^2$ increasing from violet to dark red and atomic crystals from a large database of crystalline materials (gray filled squares). (b) (circles) Average $\overline{A^U}$ (plus and minus one standard deviation) plotted versus $pN^2$ for granular packings in periodic boundary conditions. $N$ ranges from $16$ to $2048$ and $p$ ranges from $10^{-7}$ to $10^{-2}$.  We also show the average $\overline{A^U}$ plus one standard deviation for atomic crystals from a large database of crystalline materials (black line with grey filled region).}
\label{fig:suppfig_au}
\end{figure}

\section{Definitions of strain matrices in Voigt notation}
\label{si:strains}

To determine all distinct elements of $C_{mn}$, we apply each of the six elementary strains $\gamma_n$ along direction $n$, expressed in Voigt notation. The first three strain matrices correspond to uniaxial compression along the $x-$, $y-$, and $z-$axes:
\noindent $\mathbf{\Gamma_{1}}=\begin{bmatrix}
-\gamma & 0 & 0\\
0 & 0 & 0 \\
0 & 0 & 0
\end{bmatrix}$, $\mathbf{\Gamma_{2}}=\begin{bmatrix}
0 & 0 & 0\\
0 & -\gamma & 0 \\
0 & 0 & 0
\end{bmatrix}$, $\mathbf{\Gamma_{3}}=\begin{bmatrix}
0 & 0 & 0\\
0 & 0 & 0\\
0 & 0 & -\gamma
\end{bmatrix}$. 

The remaining three strains represent shear strains in the $y$-$z$, $x$-$z$, and $y$-$z$ planes: 

\noindent $\mathbf{\Gamma_{4}}=\begin{bmatrix}
0 & 0 & 0\\
0 & 0 & \gamma \\
0 & 0 & 0
\end{bmatrix}$, $\mathbf{\Gamma_{5}}=\begin{bmatrix}
0 & 0 & \gamma\\
0 & 0 & 0 \\
0 & 0 & 0
\end{bmatrix}$
$\mathbf{\Gamma_{6}}=\begin{bmatrix}
0 & \gamma & 0\\
0 & 0 & 0 \\
0 & 0 & 0
\end{bmatrix}$. 
As described in the main text, we apply these strains to the grain and boundary vertex positions, i.e. ${\vec b}_\gamma=\vec{b} \,(\mathbf{\Gamma+ \mathbf{I}})$ and ${\vec r}_\gamma=\vec{r} \,(\mathbf{\Gamma+ \mathbf{I}})$, where $\mathbf{I}$ is the identity matrix, ${\vec r}_\gamma$ and ${\vec r}$ are the strained and unstrained grain positions, and ${\vec b}_\gamma$ and ${\vec b}$ are the strained and unstrained boundary vertex positions.

\section{Coordinate-invariant measures of elastic anisotropy}
\label{si:au}

In three dimensions, there are up to $21$ independent elements of the elastic modulus tensor. In principle, these elements can be combined to generate a coordinate-invariant set of $21$ values that capture the elastic response of an anisotropic solid. A coordinate-invariant description of the six independent elements of the elastic modulus tensor has been developed in two dimensions, using representation theory and harmonic decomposition~\cite{forte_unified_2014, auffray_invariant-based_2016, vallee_structure_2013}. However, an analogous coordinate-invariant decomposition in three dimensions is much more complex. 

Given the elastic modulus tensor of a material, what is the symmetry class of the material, or equivalently, how many independent parameters does the tensor possess? Refs. \cite{abramian_recovering_2020, desmorat_generic_2019} have developed an algorithm that can determine the symmetry class and normal form of a given elastic modulus tensor. In this way, the symmetry class itself is a coordinate-invariant measure of elastic anisotropy. However, the symmetry class is not sensitive to the relative amounts of elastic anisotropy. For example, a cubic elastic modulus tensor with a very small triclinic component is equivalent to a ``fully'' triclinic elastic modulus tensor. Other methods have attempted to hierarchically decompose an elastic modulus tensor into the ``best approximations'' for each symmetry class, allowing quantitative comparisons between the relative cubic, hexagonal, and triclinic strengths of a material~\cite{browaeys_decomposition_2004}. However, this decomposition method depends strongly on the orientation of the elastic modulus tensor, which makes it difficult to apply this approach to granular packings, where there is no ``proper'' frame in which to calculate the elastic modulus tensor.

Given the difficulty of using a fully coordinate-invariant approach, another method for quantifying elastic anisotropy is to calculate physically relevant responses directly. In Refs.~\cite{goodrich_jamming_2014, ranganathan_universal_2008, hendrix_self-consistent_1998, kube_elastic_2016}, the authors calculate an orientation-average of selected combinations of elastic moduli. We described a similar approach in Sec.~\ref{sec:methods}, where different physically relevant strains ${\bf \Gamma}$ (or stresses ${\bf \Sigma}$) are applied, and the response to that strain $R_{\Gamma}$ is calculated directly from the elastic modulus tensor as in Eq.~\ref{eqn:R_gamma}. $R_\Gamma$ is then integrated over all angles to determine the average value and the variance about the average as in Eqs.~\ref{eqn: integral} and~\ref{eqn: integral_ac}. Another commonly used coordinate independent measure of anisotropy is the universal anisotropy index $A^U$~\cite{ranganathan_universal_2008, kube_elastic_2016}. Using the notation introduced in Sec.~\ref{sec:methods}, the universal anisotropy index is
\begin{equation}
\label{eqn: au}
A^U = 5 \frac{G_{DC}({\bf C})}{G_{DC}({\bf S})}+ \frac{B_{DC}({\bf C})}{B_{DC}({\bf S})} -6,
\end{equation}
where ${\bf S}={\bf C}^{-1}$ is the compliance tensor and $B_{DC}$ is the average response to a uniform compressive strain,
\begin{equation}
{\bf \Gamma}_C = \begin{pmatrix}
1-\gamma & 0 &  0\\
0 & 1-\gamma & 0 \\
0 &  0 & 1-\gamma
\end{pmatrix}.
\end{equation}
In Fig. \ref{fig:suppfig_au}, we show that $A^U$ behaves qualitatively similarly to $A_G$ and $A_C$ since $A^U$ contains information about both shear and compressive anisotropy. In Fig. \ref{fig:suppfig_au} (a), we plot $A^U$ versus the shear anisotropy $A_G$ for granular packings over a range of $pN^2$ and for atomic crystals~\cite{de_jong_charting_2015}. Similar to $A_G$ and $A_C$, anisotropic granular systems at low $pN^2$ possess large values of $A^U$, while increasing $pN^2$ yields decreasing values of $A^U$.  In Fig.~\ref{fig:suppfig_au} (b), we show the ensemble-averaged $\overline{A^U}$ for granular packings as a function of $pN^2$. Similar to the behavior for $\overline{A}_G$ and $\overline{A}_C$, $\overline{A^U}$ collapses with $pN^2$ and decreases with increasing $pN^2$, reaching values that are comparable to the elastic anisotropy of crystalline materials for $pN^2 > 1$. However, since $G_{DC}({\bf S})$ and $B_{DC}({\bf S})$ can vanish, $\overline{A^U}$ can possess large fluctuations when calculating the ensemble average.

\section{Crystalline spring networks}
\label{si:spring_nets}
\begin{figure}[!htb]
\centering
\includegraphics[width = 1.0\columnwidth]{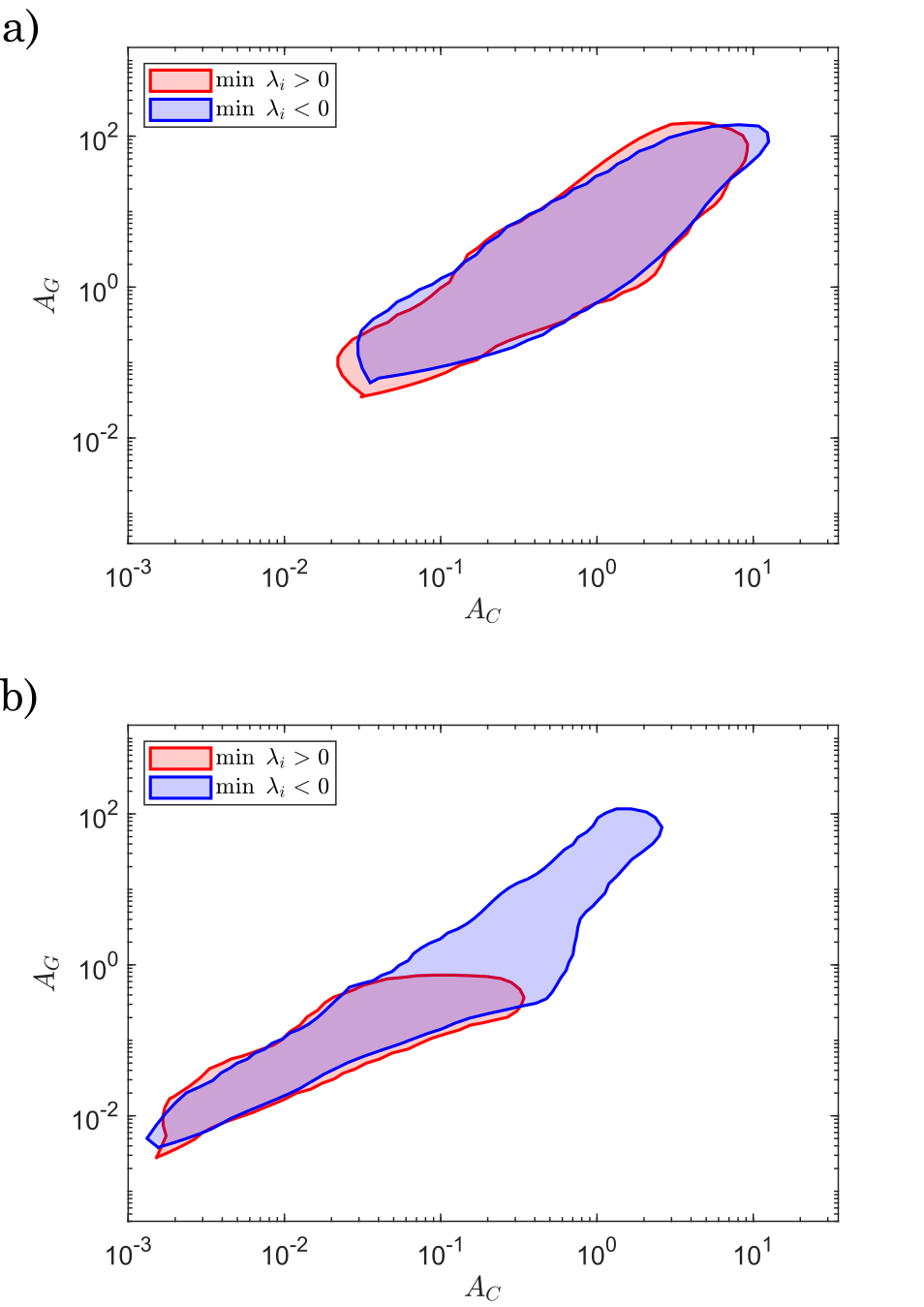}
\caption{
(a) Probability distribution $P(A^C,A^G)$ for $P\geq10^{-4}$ for granular packings with $pN^2=1$. Two colored regions are shown, one corresponding to elastic modulus tensors that are mechanically stable with all positive eigenvalues of ${\bf C}$, $c_i>0$ (red) and boundary stabilized elastic modulus tensors that possess several negative eigenvalues $c_i<0$ (blue). 
(b) Similar to (a), except for $10^6$ randomly generated strongly cubic elastic modulus tensors with $\delta=10^{-5}$.}
\label{fig:suppfig_stability}
\end{figure}
To make comparisons between the elastic anisotropy for granular packings and cyrstalline materials (whose elastic constants were obtained using DFT calculations), we also investigate the elastic anisotropy of linear spring networks. The database in Ref.~\cite{de_jong_charting_2015} contains the unit cell information (atomic positions, atomic types, and lattice parameters) for each crystalline compound. We employ a $2 \times 2 \times 2$ supercell and use the CrystalNN method in Ref.~\cite{pan_benchmarking_2021} to determine probable bonds between atoms. This method determines the bond probability between a central atom and its neighboring atoms using a function that depends on the solid angle of the Voronoi polyhedron containing the neighboring atom and atom-level properties such as the electro-negativity difference between the central and neighboring atoms. We repeat this procedure for $448$ cubic crystals in the database. Bonds between atoms are described by the double-sided spring potential:
\begin{equation}
U =  \frac{\epsilon}{2} \left( 1-\frac{r_{ij}}{r_{0,ij}} \right) ^2,
\end{equation}
where $\epsilon$ is the energy scale, $r_{ij}$ is the distance between bonded atoms $i$ and $j$, and $r_{0,ij}$ is the initial equilibrium distance between bonded atoms $i$ and $j$. We then carry out the calculations in Sec. \ref{sec:methods} to determine the elastic modulus tensor, and varying the pressure from $p=10^{-7}$ to $10^{-3}$. In the absence of defects, crystalline spring networks do not rearrange with increasing pressure, which is the primary mechanism for decreases in anisotropy in granular packings with increasing $pN^2$. Thus, the crystalline spring networks exhibit constant $A^G$ and $A^C$ for all values of $p$. Note also that increasing $N$ by adding more copies of the unit cell does not change the elastic response, and thus the crystalline spring networks are not sensitive to increasing $pN^2$. Moreover, the elastic anisotropy for these spring networks are comparable to the values obtained from DFT calculations, which indicates that the small values of elastic anisotropy arise from structural symmetry, independent of the interatomic potential (blue squares in Figs.~\ref{fig:aniso_plane} and~\ref{fig:suppfig_au}). 

\section{Mechanical stability of randomly generated elastic modulus tensors}
\label{si:stability}

To randomly generate elastic modulus tensors, we separately generate one cubic elastic tensor ${\bf C}_{\textrm{cubic}}$ (with $3$ independent constants) and another triclinic elastic tensor ${\bf C}_{\textrm{triclinic}}$ (with $21$ independent constants). The total elastic modulus tensor is the sum of the two contributions: ${\bf C} = {\bf C}_{\textrm{cubic}} + {\bf C}_{\textrm{triclinic}}$, where $\delta={\bf C}_{\textrm{cubic}}/{\bf C}_{\textrm{triclinic}}$ controls the relative strength of the symmetric and asymmetric contributions. We normalize ${\bf C}$ such that the $L_2$ norm is unity, i.e. $\|{\bf C}\|_2= \max_{i} \sqrt{c_i {\bf C}^T {\bf C}}$, where $c_i$ are the eigenvalues of ${\bf C}$. For both elastic modulus tensors, we sample individual elements from a normal distribution with mean $0$ and variance $1$. Similar results hold when sampling from a uniform distribution, and a distribution similar to experimentally measured elastic moduli, where the average compressive modulus is positive and the average shear modulus is distributed around $0$. In this Appendix, we compare results when an extra stability condition $\min_i c_i > 0$ is and is not satisfied. This condition is equivalent to $\frac{\Delta U}{V} = C_{mn}\Gamma_{n}\Gamma_{m}$ for all $\Gamma_m$ and $\Gamma_n$.

For strain-controlled deformations, it is possible to obtain a negative shear modulus $G$, as the strain is applied directly to the material (including the boundary) and the unstable eigenmode is stabilized by the fixed boundaries. Previous work has noted that the probability of $G<0$ for granular packings is maximal at $pN^2\sim1$ in the regime with large elastic anisotropy~\cite{zhang_local_2023}. Because of this result, we include granular packings and randomly generated ${\bf C}$ matrices with negative eigenvalues in the analysis. We focus on the case of strongly cubic elastic modulus tensors with $\delta=10^{-5}$. We generate random ${\bf C}$ matrices where half have $\min_i c_i>0$ and half have $\min_i c_i<0$. In Fig.~\ref{fig:suppfig_stability} (a), we show the probability distribution $P(A_C, A_G)$ with $P\geq 10^{-4}$ for granular packings generated at $pN^2\sim1$ separated into two classes: $\min_i \lambda_i > 0$ (red) and $\min_i \lambda_i < 0$ (blue). The probability boundary is smoothed with a Savitzky-Golay filter of order $2$ and window size $9$. We find that it is possible to attain large elastic anisotropy independent of the sign of $\lambda_i$. In Fig.~\ref{fig:suppfig_stability} (b), we show the probability distribution $P(A_C, A_G)$ with $P\geq 10^{-4}$ for randomly generated ${\bf C}$ matrices. In contrast, for randomly generated elastic modulus tensors, we find that only the regions with large anisotropy $A_G\geq10^0$ and $A_C\geq10^{-1}$ are associated with ${\bf C}$ with unstable eigenvalues. 

\section{Analytical expressions for elastic anisotropy}
\label{si:analytics}

In this article, we use two normalized measures of elastic anisotropy: the shear anisotropy $G_{AC}/G_{DC}$ and compressive anisotropy $U_{AC}/(B_{DC}G_{DC})^{1/2}$. For an arbitrary triclinic elastic modulus tensor $C_{mn}$, the following formulas (using Voigt notation) can be used to calculate $G_{AC}$, $G_{DC}$, $U_{AC}$, and $B_{DC}$: 

\begin{widetext}
\begin{equation}
G_{DC} = \frac{4}{15} \left(C_{11}-C_{12}-C_{13}+C_{22}-C_{23}+C_{33}+ 3 \left(C_{44}+C_{55}+C_{66}\right)\right),
\end{equation}
\end{widetext}

\begin{widetext}
\begin{equation}
B_{DC} = \frac{1}{9} \left(C_{11}+2 C_{12}+2 C_{13}+3 C_{22}+2 C_{23}+3 C_{33}\right),
\end{equation}
\end{widetext}

\begin{widetext}
\begin{dmath}
    (G_{AC})^2 = -\frac{256 C_{11} C_{12}}{1575}-\frac{256 C_{11} C_{13}}{1575}+\frac{16 C_{11} C_{22}}{1575}+\frac{32 C_{11} C_{23}}{225}+\frac{16 C_{11} C_{33}}{1575}-\frac{64 C_{11} C_{44}}{525}-\frac{64 C_{11} C_{55}}{525}-\frac{64 C_{11} C_{66}}{525}+\frac{128 C_{11}^2}{1575}-\frac{32 C_{12} C_{13}}{225}-\frac{256 C_{12} C_{22}}{1575}-\frac{32 C_{12} C_{23}}{225}+\frac{32 C_{12} C_{33}}{225}+\frac{64 C_{12} C_{44}}{525}+\frac{64 C_{12} C_{55}}{525}+\frac{64 C_{12} C_{66}}{525}+\frac{368 C_{12}^2}{1575}+\frac{32 C_{13} C_{22}}{225}-\frac{32 C_{13} C_{23}}{225}-\frac{256 C_{13} C_{33}}{1575}+\frac{64 C_{13} C_{44}}{525}+\frac{64 C_{13} C_{55}}{525}+\frac{64 C_{13} C_{66}}{525}+\frac{368 C_{13}^2}{1575}-\frac{64 C_{14} C_{24}}{105}-\frac{64 C_{14} C_{34}}{105}+\frac{64 C_{14}^2}{105}-\frac{64 C_{15} C_{25}}{105}-\frac{64 C_{15} C_{35}}{105}+\frac{64 C_{15}^2}{105}-\frac{64 C_{16} C_{26}}{105}-\frac{64 C_{16} C_{36}}{105}+\frac{64 C_{16}^2}{105}-\frac{256 C_{22} C_{23}}{1575}+\frac{16 C_{22} C_{33}}{1575}-\frac{64 C_{22} C_{44}}{525}-\frac{64 C_{22} C_{55}}{525}-\frac{64 C_{22} C_{66}}{525}+\frac{128 C_{22}^2}{1575}-\frac{256 C_{23} C_{33}}{1575}+\frac{64 C_{23} C_{44}}{525}+\frac{64 C_{23} C_{55}}{525}+\frac{64 C_{23} C_{66}}{525}+\frac{368 C_{23}^2}{1575}-\frac{64 C_{24} C_{34}}{105}+\frac{64 C_{24}^2}{105}-\frac{64 C_{25} C_{35}}{105}+\frac{64 C_{25}^2}{105}-\frac{64 C_{26} C_{36}}{105}+\frac{64 C_{26}^2}{105}-\frac{64 C_{33} C_{44}}{525}-\frac{64 C_{33} C_{55}}{525}-\frac{64 C_{33} C_{66}}{525}+\frac{128 C_{33}^2}{1575}+\frac{64 C_{34}^2}{105}+\frac{64 C_{35}^2}{105}+\frac{64 C_{36}^2}{105}-\frac{64 C_{44} C_{55}}{175}-\frac{64 C_{44} C_{66}}{175}+\frac{128 C_{44}^2}{175}+\frac{64 C_{45}^2}{35}+\frac{64 C_{46}^2}{35}-\frac{64 C_{55} C_{66}}{175}+\frac{128 C_{55}^2}{175}+\frac{64 C_{56}^2}{35}+\frac{128 C_{66}^2}{175},
\end{dmath}
\end{widetext}
and
\begin{widetext}
\begin{dmath}
(U_{AC})^2 = \frac{16 C_{11}^2}{225}+\frac{16 C_{12} C_{11}}{1575}+\frac{16 C_{13} C_{11}}{1575}+\frac{32 C_{55} C_{11}}{1575}+\frac{32 C_{66} C_{11}}{1575}-\frac{32 C_{22} C_{11}}{525}-\frac{32 C_{33} C_{11}}{525}-\frac{64 C_{23} C_{11}}{1575}-\frac{128 C_{44} C_{11}}{1575}+\frac{32 C_{12}^2}{1575}+\frac{32 C_{13}^2}{1575}+\frac{16 C_{14}^2}{315}+\frac{16 C_{15}^2}{63}+\frac{16 C_{16}^2}{63}+\frac{16 C_{22}^2}{225}+\frac{32 C_{23}^2}{1575}+\frac{16 C_{24}^2}{63}+\frac{16 C_{25}^2}{315}+\frac{16 C_{26}^2}{63}+\frac{16 C_{33}^2}{225}+\frac{16 C_{34}^2}{63}+\frac{16 C_{35}^2}{63}+\frac{16 C_{36}^2}{315}+\frac{128 C_{44}^2}{1575}+\frac{64 C_{45}^2}{315}+\frac{64 C_{46}^2}{315}+\frac{128 C_{55}^2}{1575}+\frac{64 C_{56}^2}{315}+\frac{128 C_{66}^2}{1575}+\frac{16 C_{12} C_{22}}{1575}+\frac{16 C_{22} C_{23}}{1575}+\frac{32 C_{14} C_{24}}{315}+\frac{32 C_{15} C_{25}}{315}+\frac{32 C_{16} C_{26}}{105}+\frac{16 C_{13} C_{33}}{1575}+\frac{16 C_{23} C_{33}}{1575}+\frac{32 C_{14} C_{34}}{315}+\frac{32 C_{24} C_{34}}{105}+\frac{32 C_{15} C_{35}}{105}+\frac{32 C_{25} C_{35}}{315}+\frac{32 C_{16} C_{36}}{315}+\frac{32 C_{26} C_{36}}{315}+\frac{32 C_{22} C_{44}}{1575}+\frac{128 C_{23} C_{44}}{1575}+\frac{32 C_{33} C_{44}}{1575}+\frac{64 C_{16} C_{45}}{315}+\frac{64 C_{26} C_{45}}{315}+\frac{64 C_{36} C_{45}}{315}+\frac{64 C_{15} C_{46}}{315}+\frac{64 C_{25} C_{46}}{315}+\frac{64 C_{35} C_{46}}{315}+\frac{128 C_{13} C_{55}}{1575}+\frac{32 C_{33} C_{55}}{1575}+\frac{64 C_{14} C_{56}}{315}+\frac{64 C_{24} C_{56}}{315}+\frac{64 C_{34} C_{56}}{315}+\frac{128 C_{12} C_{66}}{1575}+\frac{32 C_{22} C_{66}}{1575}-\frac{32 C_{22} C_{33}}{525}-\frac{16 C_{12} C_{13}}{1575}-\frac{64 C_{13} C_{22}}{1575}-\frac{16 C_{12} C_{23}}{1575}-\frac{16 C_{13} C_{23}}{1575}-\frac{64 C_{12} C_{33}}{1575}-\frac{32 C_{12} C_{44}}{1575}-\frac{32 C_{13} C_{44}}{1575}-\frac{32 C_{12} C_{55}}{1575}-\frac{128 C_{22} C_{55}}{1575}-\frac{32 C_{23} C_{55}}{1575}-\frac{64 C_{44} C_{55}}{1575}-\frac{32 C_{13} C_{66}}{1575}-\frac{32 C_{23} C_{66}}{1575}-\frac{128 C_{33} C_{66}}{1575}-\frac{64 C_{44} C_{66}}{1575}-\frac{64 C_{55} C_{66}}{1575}.
\end{dmath}
\end{widetext}
These expressions were obtained from analytical evaluation of the integrals in Eqs.~\ref{eqn: integral} and~\ref{eqn: integral_ac} using the tensor manipulation package OGRe in Mathematica~\cite{shoshany_ogre_2021}. Note that $G_{DC}$ and $B_{DC}$ are linear functions of $C_{mn}$, while $G_{AC}$ and $U_{AC}$ are quadratic functions of $C_{mn}$.

\end{document}